\documentclass[%
11pt,
onecolumn,
tightenlines,
superscriptaddress,
notitlepage,
preprintnumbers,
nofootinbib,
amsmath,amssymb,amsthm,
aps,
eqsecnum
]{revtex4-2}

\usepackage{graphicx}
\usepackage{endnotes}
\usepackage{setspace}
\usepackage{times}
\usepackage{helvet}
\usepackage{courier}
\usepackage{url}
\usepackage{dcolumn}

\usepackage[english]{babel}
\usepackage[utf8]{inputenc}
\usepackage[pdftex, pdftitle={Article}, pdfauthor={Author}]{hyperref}
\usepackage{fancyhdr}
\usepackage{isomath}
\usepackage{amsmath}
\usepackage{amsbsy}
\usepackage{amssymb}
\usepackage{amsthm}
\usepackage{amscd}
\usepackage{amsfonts}
\usepackage{graphics}
\usepackage{verbatim}
\usepackage{subfigure}
\usepackage{xspace}
\usepackage{euscript}
\usepackage{alltt}
\usepackage{boxedminipage}
\usepackage{float}
\usepackage{color}
\usepackage[all]{xy}
\usepackage{t1enc}
\usepackage{exscale}
\usepackage{graphicx,calc}
\usepackage{mdwlist}
\usepackage{stmaryrd}
\usepackage{units}
\usepackage{setspace}
\usepackage[english]{babel}
\usepackage[utf8]{inputenc}
\usepackage{algorithm}
\usepackage{mathtools}
\usepackage{mathrsfs}
\usepackage{bm}
\usepackage{paralist}
\usepackage[plain]{fancyref}
\usepackage{multirow}

\usepackage{ulem}

\newcommand{\dihedral}[1]{\mathcal{D}_{#1}}

\newcommand{\reflOp}{\bm{\sigma}}
\newcommand{\refl}[1]{\reflOp_{#1}}

\newcommand{\SOThree}{SO\left(3\right)}

\newcommand{\bvec}{\hat{\mathbf{e}}}
\newcommand{\eone}{\bvec_1}
\newcommand{\etwo}{\bvec_2}
\newcommand{\ethree}{\bvec_3}
\newcommand{\rotMatrix}{\mathbf{Q}}
\newcommand{\rotAbout}[1]{\rotMatrix_{#1}}
\newcommand{\rotate}[2]{\rotAbout{#1}\left(#2\right)}

\newcommand{\unitVec}{\hat{\mathbf{u}}}

\DeclareMathOperator{\Orb}{Orb}

\newcommand{\orbit}[2]{\Orb_{#2} #1}
\newcommand{\group}{\mathcal{G}}

\newcommand{\genBy}[1]{\left\langle #1 \right\rangle}

\newcommand{\NCreases}{N}
\newcommand{\creaseVec}{\hat{\mathbf{c}}}
\newcommand{\CreaseVec}{\hat{\mathbf{b}}}
\newcommand{\foldAngle}{\phi}
\newcommand{\partFoldAngle}{\rho}

\newcommand{\rk}{r_k}
\newcommand{\defMap}{\bm{\Phi}}

\newcommand{\F}{\mathbf{F}}
\newcommand{\facet}[1]{\cBody^{\left(#1\right)}}
\newcommand{\Facet}[1]{\rBody^{\left(#1\right)}}
\newcommand{\sectAngle}{\alpha}
\newcommand{\convToFoldAngle}{f_{\foldAngle}}

\newcommand{\elevAngle}{\upsilon}

\newcommand{\azi}{\varphi}
\newcommand{\xPMag}{x}
\newcommand{\xP}{\mathbf{\xPMag}}
\newcommand{\xCMag}{y}
\newcommand{\xC}{\mathbf{\xCMag}}
\newcommand{\rad}{r}
\newcommand{\Rad}{R}
\newcommand{\genBasis}{\mathbf{e}}
\newcommand{\metric}{g}
\newcommand{\metricTens}{\mathbf{\metric}}

\newcommand{\unNormalVec}{\mathbf{n}}
\newcommand{\arealStretch}{\mu}

\newcommand{\df}[1]{\mathrm{ d}#1 \:}

\DeclareMathOperator{\Grad}{Grad}
\DeclareMathOperator{\diag}{diag}
\newcommand{\cBody}{\Omega}

\newcommand{\rBody}{\cBody_0}

\newcommand{\Reals}{\mathbb{R}}

\newcommand{\Naturals}{\mathbb{N}}

\newcommand{\genAngle}{\varphi}
\newcommand{\Iden}{\mathbf{I}}

\DeclareMathOperator{\sgn}{sgn}
\DeclareMathOperator{\cSin}{sin}
\DeclareMathOperator{\cCos}{cos}

\newcommand{\numer}{\mathcal{Y}}
\newcommand{\denom}{\mathcal{X}}
\newcommand{\argSign}{\mathcal{S}}

\newcommand{\U}{U}
\newcommand{\kCrease}{k}
\newcommand{\foldAngleO}{\foldAngle^0}

\newcommand{\bMap}{\bm{\varphi}}
\newcommand{\dphi}{\frac{\mathrm{d} \bMap}{\mathrm{d} s}}
\newcommand{\ddphi}{\frac{\mathrm{d}^2 \bMap}{\mathrm{d} s^2}}
\makeatletter
\newcommand*{\gnuplotinput}[2][1.0]{%
  \begingroup
  \let\@gnplt@input@includegraphics=\includegraphics
  \def\includegraphics##1{\@gnplt@input@includegraphics[scale=#1]{#2}}%
  \let\@gnplt@input@picture=\picture
  \def\picture{\unitlength=#1\unitlength\relax\@gnplt@input@picture}%
  \input{#2}%
  \endgroup
}
\makeatother
\graphicspath{ {figs/} }

\newcommand*{\fancyrefapplabelprefix}{app}
\frefformat{plain}{\fancyrefapplabelprefix}{appendix~#1}
\Frefformat{plain}{\fancyrefapplabelprefix}{Appendix~#1}
\frefformat{main}{\fancyrefapplabelprefix}{appendix~#1}
\Frefformat{main}{\fancyrefapplabelprefix}{Appendix~#1}

\definecolor{myorange}{rgb}{1, 0.647, 0}
\renewcommand{\emph}[1]{\textit{#1}}
\newcommand{\hl}[1]{{#1}}
\newenvironment{hlbreakable}%
{}%
{}

\begin{document}

\title{Lagrangian approach to origami vertex analysis: Multistability}

\author{Matthew Grasinger}
\email[Corresponding author: ]{matthew.grasinger.1@us.af.mil}

\author{Andrew Gillman}

\author{Philip R. Buskohl}
\email[Corresponding author: ]{philip.buskohl.1@us.af.mil}

\affiliation{Foundational Technologies Directorate, Air Force Research Laboratory, Wright-Patterson AFB}


\preprint{To appear in Proceedings of the Royal Society A: \url{https://doi.org/10.1098/rspa.2026.0429}}

\begin{abstract}
    Studying the multistability of origami structures presents challenges due to the nonlinearity of their kinematics and the high-dimensional configuration spaces that are difficult to visualize and explore exhaustively. To address this, we utilize the Lagrangian framework for origami to exploit symmetries and obtain reduced-dimensional slices of the configuration space. Our analysis of degree-6 vertices with reflection symmetry reveals topological transitions in their kinematic space as sector angles are varied, with implications for the number of symmetry-constrained minima and the emergence of metastable regions. These lower-dimensional slices are amenable to exhaustive search and visualization. \hl{A subsequent full-space stability analysis shows that 18 of 41 degree-6 and 14 of 45 degree-8 symmetry-constrained minima remain minima when all locally compatible perturbations, including those that break symmetry, are admitted. The metastable regions, which would likely be overlooked by numerical optimization alone, are influenced by the interplay between the boundaries of admissible kinematic space and crease mechanical properties.} We extend our analysis to cone-like vertices with higher symmetry and one-degree-of-freedom kinematics, exploring symmetry-breaking phenomena, combinatorial structure, and their consequences for \hl{branchwise stability}. The stability landscapes uncovered have potential applications in mechanical metamaterials, mechanical computing, origami-based robotics, and structures designed to self-deploy and retain their shape.
\end{abstract}

\maketitle
\section{Introduction}
\label{sec:intro}
Origami has transformed into a powerful design paradigm across science and engineering. Its principles are enabling innovations from large-scale deployable structures in aerospace applications~\cite{miura1985method,zirbel2013accommodating,pruett2022optimizing,fuchi2021design,zhou2025hyper} to miniature devices for biomedical applications~\cite{kuribayashi2006self,andersen2009self} and other nanotechnologies~\cite{grasingerIPnanoscale}. Central to many of these advancements is the phenomenon of multistability: the ability of an origami structure to possess multiple stable equilibrium configurations. This property is particularly valuable for creating structures stable in both compact and deployed states~\cite{melancon2021multistable,li2020theory,pruett2024characterization,deshpande2024golden,dorn2023multi,dorn2021structures}, driving locomotion in origami-based robotics~\cite{novelino2020untethered,wu2021stretchable,sadeghi2021tmp,bhovad2019peristaltic,mondal2026programmable}, storing mechanical energy and information~\cite{treml2018origami,jules2022delicate,bhovad2021physical,li2016recoverable}, and designing novel mechanical metamaterials~\cite{jamalimehr2026entangled,fang2016self,yasuda2015reentrant,liu2021novel,fang2017asymmetric,schenk2013geometry,silverberg2014using,filipov2015origami,brunck2016elastic,liu2018topological,zhai2020situ,miyazawa2021heterogeneous,wen2021stacked,karami2024curved,sun2024curved}. Understanding and harnessing origami multistability is therefore a key objective for advancing these diverse applications.

Investigating multistability often begins with the common and effective idealization of rigid-facet origami, where deformation occurs solely through folding along predefined creases, changing the dihedral angles between facets. This folding is governed by geometric compatibility constraints that ensure that the ``paper'' does not tear~\cite{hull2002modelling}. Based on these constraints, some crease patterns are inflexible (only flat), others exhibit continuous families of folded states (rigidly foldable), and some possess multiple, kinematically isolated stable states (rigidly multistable). In rigidly multistable structures, such as certain cylindrical patterns, finite deformations are required to transition between stable configurations, making them robust yet reconfigurable~\cite{feng2020helical,bos2017incompressibility,reid2017geometry,silverberg2015origami}.
While this notion of multistability is important, many origami vertices and crease patterns of interest are rigidly foldable.

The complexity of rigidly foldable multistability has been an active area of research in recent years. Seemingly simple degree-$4$ vertices, often exhibiting continuous folding paths, can achieve bistability or higher-order stability (up to six states) when crease elasticity is considered~\cite{waitukaitis2015origami}. More complex systems like Stacked Miura-Ori (SMO) and its variants demonstrate rich multistable behaviors, including self-locking and asymmetric energy barriers~\cite{li2015fluidic,fang2016self,fang2017asymmetric,liu2021novel}. Other notable examples include waterbomb patterns~\cite{hanna2014waterbomb,hanna2015force,grasinger2022multistability,zhou2023low,gillman2018truss}, the Tachi-Miura Polyhedron~\cite{yasuda2015reentrant,sadeghi2021tmp}, the Yoshimura~\cite{zhou2025hyper,deshpande2024golden}, the Kresling pattern~\cite{jianguo2015bistable,novelino2020untethered,bhovad2019peristaltic,wu2021stretchable,lu2022conical,jules2022delicate}, the hypar~\cite{demaine2011non,liu2019invariant}, and even curved crease origami~\cite{sun2024curved}, each finding applications exploiting their unique stability landscapes. 
\hl{Recent work has developed theory and design methods for the related but distinct concept of multi-configuration rigidity, in which prestress maintains contact with unilateral constraints at multiple prescribed configurations, producing contact-pinned boundary minima that can resist finite perturbation loads without displacement until contact is lost~\cite{dorn2023multi,dorn2021structures}.}
\hl{Despite these advances, characterizing multistability, broadly speaking,} presents significant challenges due to inherent geometric nonlinearities in folding kinematics and the potentially large number of degrees of freedom, especially in complex patterns, and, in some cases, non-local constraints like self-contact~\cite{liu2018topological}.

These complexities often necessitate experimental exploration or computationally intensive numerical methods (such as finite element analysis or bar-hinge models) to map the energy landscape and identify stable states~\cite{hanna2014waterbomb,hanna2015force,brunck2016elastic,fang2017asymmetric,gillman2018truss,gillman2019design,li2015fluidic,fang2016self,yasuda2015reentrant,liu2021novel,sadeghi2021tmp,li2020theory,fang2019magneto}. While powerful, experiments can be difficult to scale across large design spaces, and numerical optimization or sampling techniques may struggle to guarantee the discovery of all stable configurations or novel behaviors. This highlights a critical need for alternative, more efficient methods to analyze the folding and stability of origami patterns.

Analytical approaches offer a pathway to potentially overcome these limitations by providing explicit mathematical descriptions of the folding process. Seminal work by Huffman utilized the Gauss map and spherical trigonometry to derive implicit kinematic relationships for degree-4 vertices~\cite{huffman1976curvature}. This geometric perspective was later extended to obtain explicit solutions for specific symmetric structures like the 8-fold waterbomb vertex~\cite{hanna2014waterbomb}, demonstrating the power of analytical techniques for capturing complex folding behavior.

An alternative and powerful analytical framework stems from modeling origami folding using concepts from continuum mechanics, specifically notions of the deformation map and deformation gradient within the framework of Lagrangian kinematics~\cite{hull2002modelling,feng2020designs,feng2020helical,liu2021origami,liu2024design,grasinger2024lagrangian}. This perspective can naturally incorporate both symmetry and compatibility constraints. Early work recognized the connection between sequences of rotations along closed loops and compatibility, laying the groundwork for this approach~\cite{hull2002modelling}. Subsequent research leveraged this Lagrangian viewpoint combined with group theory and objective structures~\cite{james2006objective} to derive simplified kinematics for degree-4 vertices, generate complex folded states (e.g., helical origami~\cite{feng2020helical}, curved crease origami~\cite{liu2024design}), and derive explicit kinematic solutions for general degree-4~\cite{foschi2022explicit}, symmetric degree-6~\cite{farnham2022rigid,grasinger2024lagrangian}, and symmetric degree-8 vertices~\cite{grasinger2024lagrangian}, including non-Euclidean cases~\cite{foschi2022explicit}.

A key advantage of the Lagrangian approach is its ability to systematically incorporate geometric constraints, including symmetry. In our previous study~\cite{grasinger2024lagrangian}, we leveraged this capability to derive reduced-order compatibility conditions specifically for origami vertices possessing reflectional and/or rotational symmetries. By exploiting these symmetries, we obtained exact, lower-dimensional kinematic solutions describing the multi-degree-of-freedom folding pathways for various symmetric vertices, significantly simplifying their description compared to a full-dimensional kinematic analysis.

Building directly on these findings, this paper utilizes the exact, lower-dimensional kinematic solutions derived in~\cite{grasinger2024lagrangian} to efficiently investigate origami multistability. By formulating the energy landscape within these reduced configuration spaces defined by symmetry, we can systematically search for and characterize stable folded states. This approach allows for a targeted and computationally efficient exploration of multistability, particularly for states conforming to specific symmetries, offering a powerful complement to existing numerical and experimental techniques for navigating the complex design space of origami.

The structure of this paper is as follows: \Fref{sec:kinematics} briefly summarizes the key developments of our previous work regarding the kinematics of symmetric origami vertices.
\hl{The contributions of the present work begin with the energetic analysis in \Fref{sec:multistability}.
Specifically, we use the previously derived symmetry-reduced kinematics to \begin{inparaenum}[(1)]
   \item exhaustively map the (symmetrically folded) energy landscapes and multistability of degree-$6$ and degree-$8$ vertices, and
   \item relate multistability and metastability to the topology and boundaries of admissible kinematic space.
\end{inparaenum}}
\Fref{sec:origami-cone} explores the kinematics, symmetry breaking, and combinatorics of origami analogs of a cone with $\NCreases$ creases.
\Fref{sec:cone-multistability} considers the number of stable states of origami cones in mechanical property space.
The work concludes with \Fref{sec:conclusion}.

\section{\hl{Review of} symmetric kinematics} \label{sec:kinematics} 
\begin{figure}
	\centering
	\includegraphics[width=0.8\linewidth]{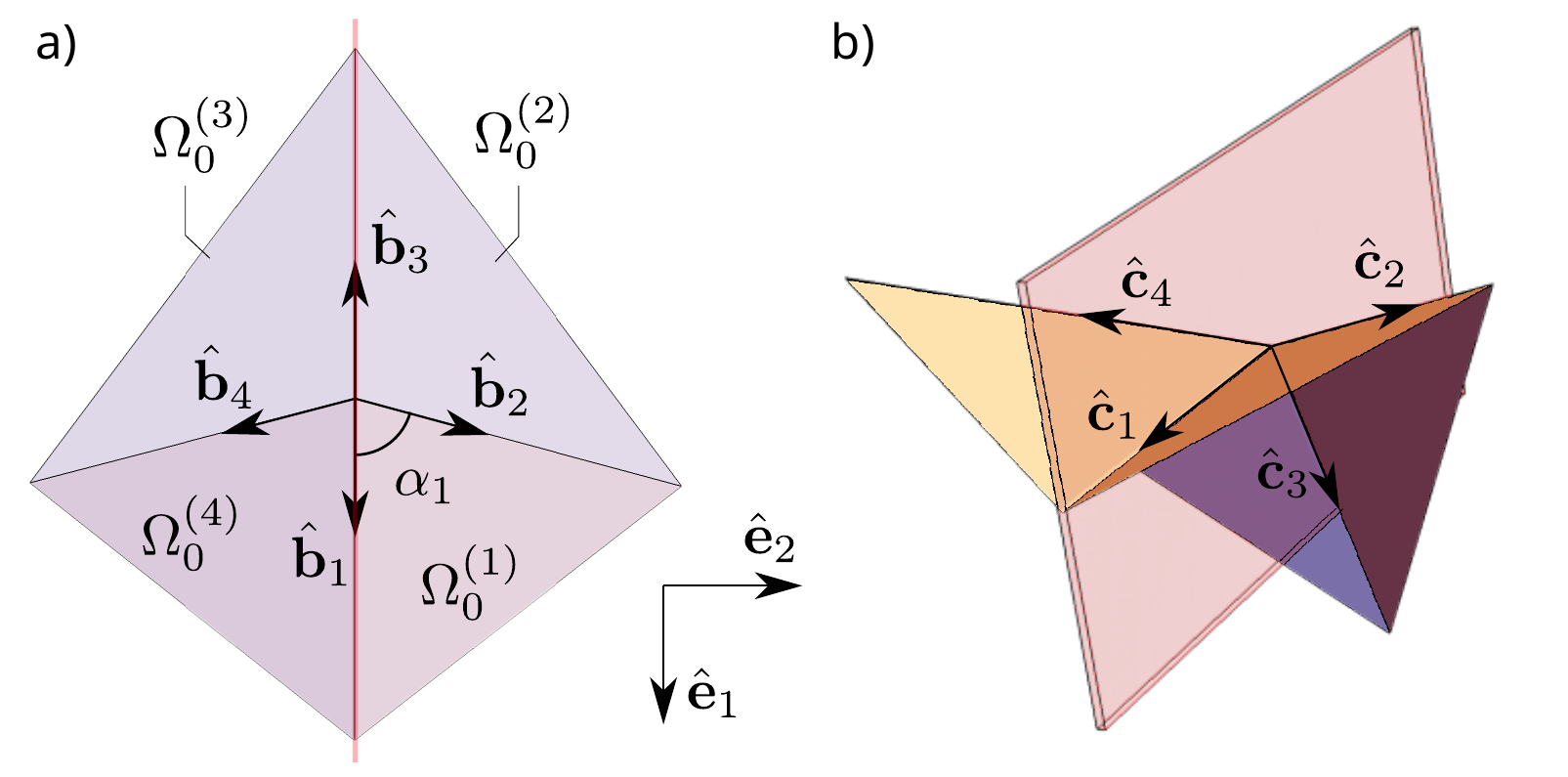}
	\caption{
		\textbf{Folding vertices with reflection symmetry.}
		\textbf{a)} an example vertex with reflection symmetry in the reference configuration. $\CreaseVec_i$ denotes the unit vector along crease $i$, $\Facet{i}$ denotes the facet bounded by creases $i$ and $i+1$.
        \textbf{b)} Schematic of the reflection condition: if the right half of the vertex can be folded such that its left edge lies in the plane of the reflection symmetry, then a compatibly folded vertex can be generated by taking a copy of the half vertex about the plane of reflection (i.e., by taking the group orbit of the underlying symmetry group).
	}
	\label{fig:reflection}
\end{figure}

\hl{The reduced compatibility conditions and symmetric kinematic results in this section were previously derived in Ref~\cite{grasinger2024lagrangian}.
They are reviewed here to establish the notation and analytical results to be used for the new multistability and energy landscape analysis developed in Sections 3-5.}
In this work, we will consider the deformation of single vertices assuming that (to a good approximation) all of the deformation occurs about the creases and the facets can be idealized as rigid.
Starting with any crease, let the unit vector which begins at the vertex and extends along the crease in the reference configuration be denoted by $\CreaseVec_1$; then proceed around the vertex in the counterclockwise direction, labeling the unit vector along each subsequent crease, $\CreaseVec_2, \CreaseVec_3, \dots, \CreaseVec_\NCreases$, where $\NCreases$ is the number of creases at the vertex.
The fold angle at $\CreaseVec_i$ is denoted by $\foldAngle_i \in \left[-\pi, \pi\right]$.
For the facets, let the facet that is bounded by the crease along $\CreaseVec_i$ and $\CreaseVec_{i+1}$ be denoted by $\Facet{i}$ and $\facet{i}$ in the reference configuration and deformed configuration, respectively, for $i = 1, \dots, \NCreases-1$; finally, $\Facet{\NCreases}$ and $\facet{\NCreases}$ is the facet bounded by the creases $\NCreases$ and $1$.
An example of these \hl{definitions} is shown in \fref{fig:reflection}.

\paragraph*{Loop closure constraint.}
Next we consider constraints on the allowable fold angles such that the origami can be folded without tears (i.e., the \emph{compatibility constraint} is satisfied).
In general, this takes the form of the loop-closure constraint, and can be derived by assuming that one of the facets is fixed (e.g., $\facet{\NCreases} = \Facet{\NCreases}$)
(to fix the origami with respect to rigid body translations and rotations\footnote{Here, by ``rigid rotation'', we mean that the entire origami is rotated such that all of the fold angles are preserved.}).
Let $\rotAbout{\unitVec}\left(\genAngle\right) \in \SOThree$ denote a rotation of angle $\genAngle$ about the axis $\unitVec$.
Then, because there are no tears between facets,
\begin{equation} \label{eq:fold-map}
	\xC = \defMap\left(\xP\right) = \left(\prod_{i=1}^{j} \rotAbout{\CreaseVec_i}\left(\foldAngle_i\right)\right) \xP, \quad \xP \in \Facet{j},
\end{equation}
As $\defMap\left(\xP\right)$ is a linear transformation of $\xP$, the deformation gradient is simply
\begin{equation}
	\F\left(\xP\right) = \Grad \defMap = \prod_{i=1}^j \rotAbout{\CreaseVec_i}\left(\foldAngle_i\right), \quad \xP \in \Facet{j},
\end{equation}
which is piecewise constant.
For brevity, let $\F_i$ be defined such that $\facet{i} = \F_i \Facet{i}$.
Recall that the above formulation of the deformation map assumes that \begin{inparaenum}[1)] \item $\Facet{\NCreases}$ is fixed and \item there are no tears (i.e., the deformation is compatible). \end{inparaenum}
As a result, a necessary condition for allowable fold angles is given by the loop-closure constraint~\cite{hull2002modelling,feng2020helical}
\begin{equation} \label{eq:compatibility}
	\F_{\NCreases}\left(\foldAngle_1, \dots, \foldAngle_{\NCreases}\right) = \prod_{i=1}^{\NCreases} \rotAbout{\CreaseVec_i}\left(\foldAngle_i\right) = \Iden.
\end{equation}
Choosing the deformation map from among rigid body translations and rotations of the origami such that $\F_{\NCreases} = \Iden$ is a standard convention that is convenient for deriving \eqref{eq:compatibility}.
However, we will adjust this convention on the basis of the symmetry of the underlying crease pattern in order to derive the reduced-order compatibility conditions.

\hl{
\paragraph*{Scope of symmetry reductions.}
Having introduced the general loop-closure constraint, we next exploit symmetries of the crease pattern to reduce the kinematic description.
The crease pattern local to a single origami vertex consists of a finite set of crease directions emanating from a common point.
Consequently, its finite planar point-symmetry group is necessarily cyclic, $C_n$, or dihedral, $\dihedral{n}$ (the dual of its line representation is a polygon).
Here we focus on reflection symmetry ($\dihedral{1}$) and a family of combined reflection-rotation symmetries ($\dihedral{2h}$), for which the Lagrangian formulation yields particularly useful reduced compatibility conditions.
These cases do not exhaust all possible symmetries; for example, purely rotational $C_n$, odd dihedral symmetries ($\dihedral{2h+1}$), and dihedral geometries outside the specific reflection-axis construction considered below are not treated explicitly herein.
The selected symmetries nevertheless arise naturally in familiar origami patterns.
A standard Miura-ori vertex is a degree $4$ vertex with reflection symmetry (although its single kinematic degree of freedom leaves no additional dimension to eliminate).
The Yoshimura pattern is constructed from degree-$6$ vertices with mirror symmetry and therefore is a classical example of $\dihedral{1}$ symmetry considered below.
Triangular Resch patterns contain both threefold-symmetric and reflection-symmetric degree-$6$ vertices, whereas the existence of global symmetries depends on their tiling.
More general symmetries become relevant for multi-vertex crease patterns; for example, helical groups have been used to construct Miura origami cylinders~\cite{feng2020helical}, and Euclidean and conformal group-orbit constructions have been developed for origami structures with curved panels~\cite{liu2024design}.
The present focus on $\dihedral{1}$ and $\dihedral{2h}$ symmetric vertices is intended as a building block toward such multi-vertex problems, whose number of degrees of freedom can grow rapidly.
}

\paragraph*{Vertices with a reflection symmetry ($\dihedral{1}$ symmetry)\footnote{By $\dihedral{n}$, we mean the dihedral group of $n$ rotational symmetries and $n$ reflection symmetries; that is, the symmetries of the regular $n$-gon.}.}
When considering vertices with an even number of creases, $\NCreases$, and a reflection symmetry about the plane which cuts through creases $1$ and $\NCreases / 2 + 1$, it will be convenient to adopt the convention
\begin{equation}
	\begin{split}
		\F_1 = \rotate{\CreaseVec_1}{\frac{\foldAngle_1}{2}}, \quad \F_2 = \rotate{\CreaseVec_1}{\frac{\foldAngle_1}{2}} \rotate{\CreaseVec_2}{\foldAngle_2}, \quad \dots, \\
		\quad \F_{\NCreases} = \rotate{\CreaseVec_1}{\frac{\foldAngle_1}{2}} \prod_{i=2}^{\NCreases} \rotate{\CreaseVec_{i}}{\foldAngle_{i}} = \rotate{\CreaseVec_1}{-\frac{\foldAngle_1}{2}}
	\end{split}
\end{equation}
which corresponds to taking the previous convention and rigidly rotating the body by $\rotate{\CreaseVec_1}{-\foldAngle_1 / 2}$, where the last equality corresponds to multiplying both sides of \eqref{eq:compatibility} by $\rotate{\CreaseVec_1}{-\foldAngle_1 / 2}$.
Let $\creaseVec_i = \F_{i-1} \CreaseVec_i$ be the unit vector along the crease $i$ in the deformed configuration.
Then, using the modified convention for the deformation map, a reduced-order compatibility condition can be formulated as~\cite{grasinger2024lagrangian}
\begin{equation} \label{eq:reflection-condition}
    \creaseVec_{\NCreases / 2 + 1} \cdot \etwo = 0.
\end{equation}
\Fref{fig:reflection} shows a geometric interpretation of this reduced-order condition: the first half of the vertex is folded such that $\creaseVec_1$ and $\creaseVec_{\NCreases/2+1}$ again lie in the same plane of reflection symmetry as the flat configuration.
By folding half of the vertex such that its interior edge lies in the plane of symmetry, the second half of the vertex can be generated by reflecting about the plane of symmetry.
Then,
\begin{subequations}
\begin{align}
\foldAngle_{\NCreases/2+2} &= \foldAngle_{\NCreases / 2}, \quad \foldAngle_{\NCreases/2+3} = \foldAngle_{\NCreases / 2 - 1}, \quad \dots, \quad \foldAngle_{\NCreases} = \foldAngle_2, \\
\creaseVec_{\NCreases/2+2} &= \refl{\etwo} \creaseVec_{\NCreases / 2}, \quad \creaseVec_{\NCreases/2+3} = \refl{\etwo} \creaseVec_{\NCreases / 2 - 1}, \quad \dots, \quad \creaseVec_{\NCreases} = \refl{\etwo} \creaseVec_2 
\end{align}
\end{subequations}
where $\refl{\unitVec} = \Iden - 2 \unitVec \otimes \unitVec$ is a reflection about the plane normal to $\unitVec$.
As a result, the number of unknowns has been reduced from $\NCreases - 3$ to $\NCreases / 2 - 1$~\cite{grasinger2024lagrangian}.

\begin{figure}
	\centering
	\includegraphics[width=0.85\linewidth]{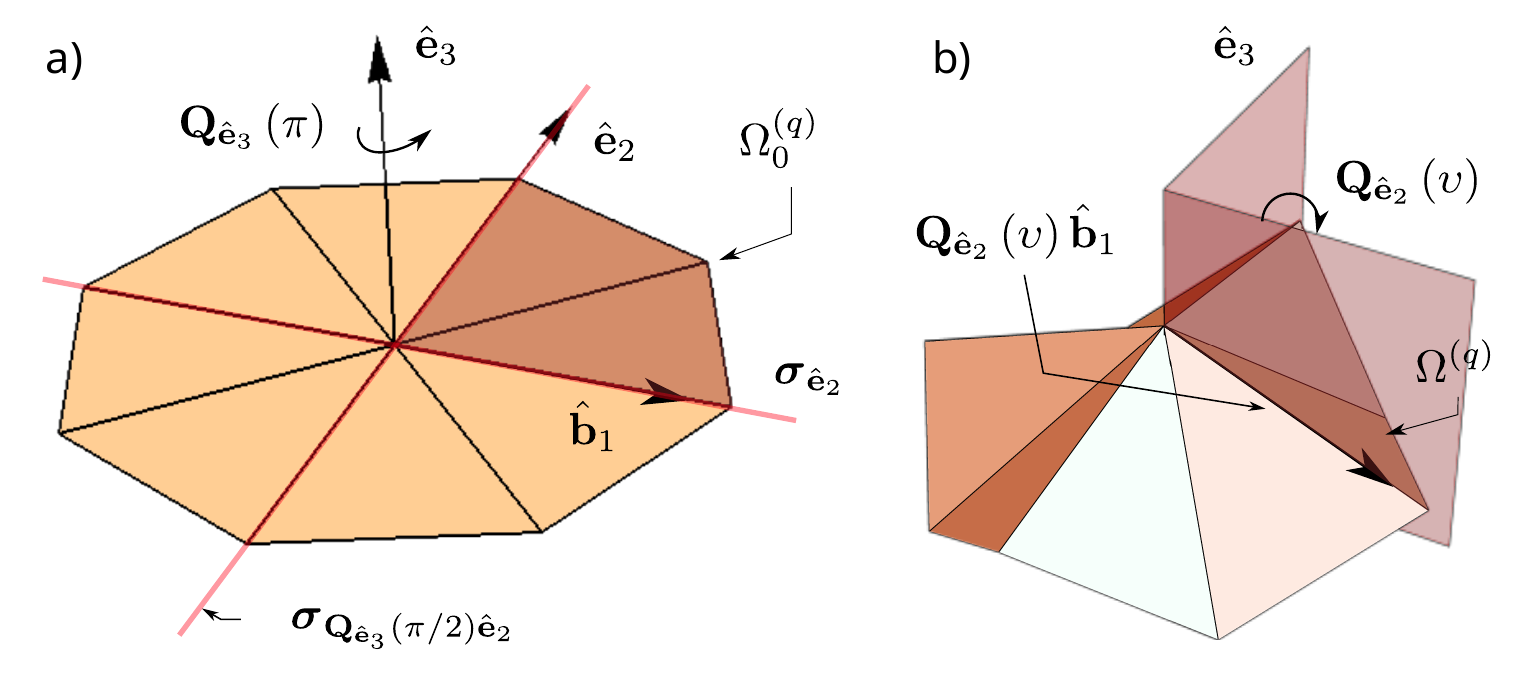}
	\caption{
		\textbf{Folding vertices with $\dihedral{2h}$ symmetry.}
		A unit cell, or ``wedge'', is elevated by angle $\elevAngle$ and then folded such that its edges are within the planes of reflection from the flat state. A compatibly folded vertex is generated by taking successive copies of reflections of the folded wedge (i.e., taking the group orbit of $\genBy{\rotate{\ethree}{\genAngle}, \refl{\etwo}}$).
	}
	\label{fig:rotation-reflection}
\end{figure}

\paragraph*{Vertices with reflection-rotation symmetries ($\dihedral{2h}, h \in \Naturals$ symmetry).} \label{sec:rotation-reflection-kinematics}
Next we consider vertices with more symmetry, both rotation and reflection symmetries.
The symmetry of interest is characterized by the group $\group = \genBy{\refl{\etwo}, \rotate{\ethree}{\genAngle}}$ where $\genBy{\Box}$ means the group generated by $\Box$.
Assume that the number of creases $\NCreases$ is even, $\genAngle = \pi / h, h \in \Naturals$ (which implies $\group \cong \dihedral{2h}$), and that there are creases in the planes of reflection (in the flat configuration).
Let $k$ be such that $\refl{\CreaseVec_k \times \ethree} \in \group$ and $\CreaseVec_{k} = \rotate{\ethree}{\genAngle/2} \CreaseVec_1$.
Then we can leverage the symmetry of the vertex to formulate a reduced-order compatibility condition by again adjusting the convention for the deformation map.
Here the key idea is to first elevate crease $1$ by some angle, $\elevAngle$, via the rotation $\rotate{\etwo}{\elevAngle}$, and then follow the convention for the reflection deformation:
\begin{equation}
		\F_1 = \rotate{\etwo}{\elevAngle} \rotate{\CreaseVec_1}{\frac{\foldAngle_1}{2}}, \quad \F_{j} = \rotate{\etwo}{\elevAngle} \rotate{\CreaseVec_1}{\frac{\foldAngle_1}{2}} \prod_{i=2}^{j} \rotate{\CreaseVec_{i}}{\foldAngle_{i}} \text{ for } j = 2, \dots, \NCreases.
\end{equation}
This is motivated by the observation that the folding of certain origami vertices resembles the center vertex ``popping'' up or down~\cite{hanna2014waterbomb,gillman2018truss,treml2018origami,grasinger2022multistability}, which is now parameterized by $\elevAngle$.
Then if one folds the fraction of the vertex bounded by creases $1$ and $k$ (i.e., the unit cell given by $\Facet{c} = \bigcup_{i=1}^{k-1} \Facet{i}$) such that creases $1$ and $k$ still remain in their corresponding planes of reflection from the flat configuration, a compatibly folded vertex can be generated, with the symmetry of the flat configuration, by taking the group orbit of the unit cell (i.e., $\bigcup_{g \in \group} \left(g \cdot \Facet{c}\right)$ where $g \cdot \Facet{c}$ is the group action of $g$ on $\Facet{c}$~\cite{grasinger2024lagrangian}).
An example of an $8$-vertex with this symmetry and the convention for the deformation map are shown in \fref{fig:rotation-reflection} a) and b), respectively.
Formally, this reduced-order compatibility is given by~\cite{grasinger2024lagrangian}
\begin{equation} \label{eq:rotation-reflection-condition}
    \frac{\creaseVec_k \cdot \etwo}{\creaseVec_k \cdot \eone} = \frac{\CreaseVec_k \cdot \etwo}{\CreaseVec_k \cdot \eone}.
\end{equation}
After taking the group orbit, the resulting fold angles are given by
\begin{equation} \label{eq:general-reflection-rotation-angle-assignments}
	\foldAngle_{k+1} = \foldAngle_{k-1}, \: \foldAngle_{k+2} = \foldAngle_{k-2}, \: \dots, \: \foldAngle_{2k-1} = \foldAngle_1, \: \foldAngle_{2 k} = \foldAngle_2, \: \foldAngle_{2 k + 1} = \foldAngle_3, \: \dots, \: \text{etc.};
\end{equation}
in other words, going counterclockwise around the vertex from crease $k+1$ to $\NCreases$, the subsequent fold angles are equivalent to the fold angle at the next crease of decreasing index until $1$, then increasing in index until $k$, decreasing until $1$, etc., until all fold angles have been specified.
As a result of this process, the number of unknowns has been reduced from $\NCreases - 3$ to $\NCreases / m$ where $m$ is the smallest integer such that $\rotate{\ethree}{m \genAngle} = \Iden$~\cite{grasinger2024lagrangian}.
\hl{Since exhaustive grid-based searches of an energy landscape suffer from the ``curse of dimensionality'', symmetry reduction can substantially reduce the computational expense associated with mapping symmetrically folded local minima.
For example, a multistart gradient-based optimization initialized on a uniform grid requires $q^d$ separate optimization runs, where $q$ is the number of starting points sampled along each of the $d$ independent coordinate directions (i.e., degrees of freedom).
Reducing the kinematic dimension therefore reduces the number of optimization runs by a factor of $q^{\NCreases - 3 - d_{\mathrm{sym}}}$ for a single vertex where $d_{\rm sym}$ is the symmetry-reduced number of degrees of freedom.
For the degree $6$ and degree-$8$ vertices considered here, this corresponds to reducing the search from $\mathcal{O}\left(q^3\right)$ to $\mathcal{O}\left(q^2\right)$ and from $\mathcal{O}\left(q^5\right)$ to $\mathcal{O}\left(q^2\right)$, respectively.
This advantage becomes increasingly significant for vertices with many creases and suggests a pathway toward more tractable exploration of multi-vertex crease patterns with otherwise prohibitively large configuration spaces (provided the structure has symmetry and symmetrically folded stable states are of interest).}

\section{Multistability of multi-degree-of-freedom vertices} \label{sec:multistability}
A solid body achieves static equilibrium when the resultant force and the resultant torque acting upon it both vanish, ensuring no linear or angular acceleration. 
This condition implies that the gradient of the potential energy vanishes, or, if constrained, that the gradient is normal to the boundary of those constraints. 
Furthermore, this static equilibrium is deemed stable if any minor disturbance from this state leads to forces and torques that act to return the body to its original equilibrium position, a condition that corresponds to the potential energy being at a local minimum.
Origami with rigid facets can achieve multistability through torsional elasticity at its creases.
Assuming the torsional elasticity is linear, the energy of the origami is
\begin{equation} \label{eq:energy}
	\U = \sum_{i=1}^\NCreases \frac{\kCrease_i}{2} \left(\foldAngle_i - \foldAngleO_i\right)^2,
\end{equation}
where $\kCrease_i$ and $\foldAngleO_i, i = 1, \dots, \NCreases$ are the torsional stiffnesses and rest fold angles of the creases, respectively.
For deployable structures, it is sometimes advantageous to design the mechanical properties of the creases such that the ``stowed'' and ``deployed'' configurations are stable states~\cite{addis2023connecting,pruett2024characterization}.
Multistability is also an important property for energy storage, energy dissipation, storing mechanical information, and mechanical computing.
Designing for multistability, however, can be nontrivial~\cite{li2020theory,grasinger2022multistability,addis2023connecting,gillman2018truss,hanna2014waterbomb,hanna2015force,sun2024curved,fang2017asymmetric,dorn2021structures,dorn2023multi}.
An efficient way to map out the energy landscape of an origami structure would be advantageous for the design of multistability.

\hl{
\paragraph*{Full-space equilibrium and stability.}
Because the crease energy and loop-closure equations are invariant under the imposed symmetry, every regular interior stationary point of the symmetry-restricted energy is also a stationary point on the full compatible manifold.
Likewise, a state pinned against a genuine symmetry-invariant self-contact constraint with nonnegative associated contact multipliers satisfies the first-order equilibrium conditions in the full contact-admissible space.
Thus, except for states located on boundaries created only by the symmetry-reduced compatibility parameterization, the remaining question is whether these equilibria remain stable to symmetry-breaking perturbations.

Accordingly, each local minimum identified in the symmetry-reduced energy landscapes is tested against all locally compatible fold-angle perturbations, including perturbations that break the imposed symmetry.
At a candidate state, the deformed crease directions are assembled into the linearized loop-closure operator, whose nullspace defines the tangent space of infinitesimally compatible fold-angle variations~\cite{tachi2009simulation}.
Finite perturbations within this tangent space are retracted onto exact nonlinear loop closure, and the gradient and Hessian of the crease energy restricted to the resulting local compatibility manifold are numerically approximated.
A candidate with negligible restricted gradient and a positive-definite restricted Hessian is classified as a strict minimum in the full compatible space.
A stationary candidate with a negative-curvature mode that remains non-self-intersecting is classified as an unstable equilibrium.

Self-contact is not imposed in this auxiliary calculation.
Consequently, a candidate that remains stable when facet penetration is permitted is necessarily stable when nonpenetration constraints are restored, whereas a computed destabilizing mode may be blocked by self-contact.
Such contact-dependent cases, together with states on boundaries arising only from the symmetry-reduced compatibility parameterization, are classified as inconclusive.
Further mathematical and numerical details are provided in Appendix~\ref{app:full-space-stability}, and the source code and state data are provided in the accompanying repository.
}

\hl{\paragraph*{Summary of model assumptions.} Before considering specific vertices, we summarize the assumptions underlying the symmetry-reduced energy landscapes considered herein:
\begin{enumerate}[(a)]
\item the facets of the origami are assumed rigid, such that deformation is localized to rotation about the creases;
\item allowable fold angles satisfy the loop-closure compatibility constraint; within the symmetry-reduced spaces considered here, compatibility is enforced through \eqref{eq:reflection-condition} for vertices with reflection symmetry (i.e., $\dihedral{1}$) and \eqref{eq:rotation-reflection-condition} for vertices with rotation and reflection symmetries (i.e., $\dihedral{2h}$);
\item the reference crease geometry and mechanical properties (specifically, the crease stiffnesses and stress-free fold angles) share the symmetry of interest;
\item the folded configurations considered are restricted to those preserving the same symmetry as the crease geometry and mechanical properties;
\item self-intersecting configurations are excluded from the admissible kinematic space.
\end{enumerate}
Thus, the energy landscapes considered below represent the contact-admissible, symmetry-preserving subset of the full compatible configuration space.}

\subsection{Multistability of degree-$6$ vertices with reflection symmetry ($\dihedral{1}$ symmetry)} \label{sec:multistability-6}

Leveraging our analytical representation of the kinematics, we map out the energy landscapes for degree-$6$ vertices with reflection symmetry: $\CreaseVec_1 = -\CreaseVec_4 = \eone$, $\CreaseVec_5 = \refl{\etwo} \CreaseVec_3$, and $\CreaseVec_6 = \refl{\etwo} \CreaseVec_2$.
We investigate how the properties of the lower-dimensional slice of the energy landscape, taken by enforcing symmetry, vary as both the geometry and torsional stiffnesses are varied.
Let $\sectAngle_i$ denote the angle between $\CreaseVec_i$ and $\CreaseVec_{i+1}$.
For simplicity, we restrict our attention to vertex geometries where $\sectAngle_3 = \sectAngle_1$, $\sectAngle_2 = \pi - 2 \sectAngle_1$, and only $\sectAngle_1$ varies independently.
For such vertices, the solution to the reduced-order compatibility condition, \eqref{eq:reflection-condition}, is~\cite{grasinger2024lagrangian}
\begin{equation} \label{eq:phi1-6-fold}
	\begin{split}
		\foldAngle_1 &= \convToFoldAngle\left(2 \arctan\left(-\argSign_6 \numer_6, \argSign_6 \denom_6\right)\right), \\
		\numer_6 &= \cCos \sectAngle_1 \cSin \sectAngle_1 \Bigg(
		2 \cCos^2 \sectAngle_1 \cSin^2 \frac{\foldAngle_3}{2} \left(1 - 2 \cCos \foldAngle_2\right) - \cSin^2 \sectAngle_1 + \\
		& \qquad \qquad \qquad \qquad \quad \cCos \foldAngle_3 \left(1 - \cCos\foldAngle_2 + \cSin^2 \sectAngle_1\right) +
		\cSin \foldAngle_2 \cSin \foldAngle_3
		\Bigg), \\
		\denom_6 &= \cSin \sectAngle_1 \left(
		\cSin \foldAngle_2 +
		2 \cCos \left(2 \sectAngle_1\right) \cSin \foldAngle_2 \cSin^2 \frac{\foldAngle_3}{2} +
		\cCos \foldAngle_2 \cSin \foldAngle_3 \right), \\
            \argSign_6 &= \pm 1,
	\end{split}
\end{equation}
where $\arctan\left(y, x\right)$ considers the quadrant in which  $\left(x, y\right)$ resides, and where $\argSign_6$ helps specify the quadrant of $\left(\denom_6, \numer_6\right)$.
Of these two possible solutions, only one satisfies compatibility for a general $\foldAngle_2, \foldAngle_3,$ and $\sectAngle_1$.
Lastly,  
\begin{equation}
	\convToFoldAngle\left(\genAngle\right) = \begin{cases}
		\convToFoldAngle\left(\genAngle + 2\pi\right) & \genAngle < -\pi \\
		\convToFoldAngle\left(\genAngle - 2\pi\right) & \genAngle > \pi \\
		\genAngle & \text{otherwise}
	\end{cases},
\end{equation}
converts fold angles into the allowable range prior to contact of adjacent faces (i.e., $\left[-\pi, \pi\right]$).

The mountain-valley ($\foldAngle > 0$ are ``valley'' and $\foldAngle < 0$ are ``mountain'') assignments of the rest fold angles are chosen to correspond with flat foldability conditions of the vertex~\cite{lang2017twists}.
Let $\foldAngle_m = -\pi$ and $\foldAngle_v = \pi / 2$.
Then,
\begin{equation}
	\begin{cases}
		\foldAngleO_1 = \foldAngle_m, \: \foldAngleO_2 = \foldAngle_v, \: \foldAngleO_3 = \foldAngle_v, \: \foldAngleO_4 = \foldAngle_m, \: \foldAngleO_5 = \foldAngle_v, \: \foldAngleO_6 = \foldAngle_v, & \sectAngle_1 \leq \pi / 3 \\
		\foldAngleO_1 = \foldAngle_v, \: \foldAngleO_2 = \foldAngle_v, \: \foldAngleO_3 = \foldAngle_m, \: \foldAngleO_4 = \foldAngle_v, \: \foldAngleO_5 = \foldAngle_m, \: \foldAngleO_6 = \foldAngle_v, & \text{otherwise}
	\end{cases}.
\end{equation}
For the torsional stiffnesses, all of the creases with a mountain rest angle have stiffness $\kCrease_m = 0.025 \sqrt{\rk}$, and the ``valley'' creases have stiffness $\kCrease_v = 0.025 / \sqrt{\rk}$, where the ratio $\rk = \kCrease_m / \kCrease_v$\footnote{
    \hl{The overall scale of the crease stiffnesses is arbitrary for the present analysis; $\rk$ controls their relative stiffness, while the common stiffness scale may be chosen sufficiently small relative to the facet stiffness to maintain the rigid-facet approximation.}
} is varied.

The energy landscapes for $\alpha_1=\pi/4$, $\pi/3$, and $5\pi/12$ (left, middle, and right columns, respectively) and $\rk=10^{-2}$, $10^0$, and $10^2$ (top, middle, and bottom rows, respectively) are shown in \Fref{fig:multistability-6}.
\hl{All stars identify local minima within the $\dihedral{1}$-symmetric configuration space.
Gold stars remain strict local minima in the full compatible kinematic space, red stars become unstable equilibria when symmetry-breaking perturbations are admitted, and gray stars denote cases whose stability cannot be resolved either (a) without explicitly treating self-contact or (b) because the second derivative test is inconclusive.
The triplet
\[
    N_{\mathrm{sym}}
    \rightarrow
    N_{\mathrm{nr}}
    \rightarrow
    N_{\mathrm{full}}
\]
in the lower corner of each panel reports, respectively, the total number of symmetry-constrained minima, the number not ruled out by the full-space analysis (gold plus gray), and the number certified as full-space stable (gold).}
When applicable, metastable boundary segments are delineated by red intervals (see the figure legend and the definition below), while white space denotes configurations excluded by self-intersection.\footnote{Self-intersection can be checked for symmetrically folded structures efficiently by leveraging the deformation map, symmetry, and checking pairs of faces and edges (see ~\cite{grasinger2024lagrangian} Appendix B)}

\hl{
\paragraph*{Metastability.}
Metastable regions identified herein occur along finite segments of the boundary of the symmetry-reduced admissible space, and are characterized by a nearly flat energy profile when the energy is restricted to that boundary.
Let $\bMap : I \subset \Reals \mapsto \Gamma \subset \Reals^{\NCreases}, s \mapsto \bMap(s)$ be a sufficiently smooth arc length parameterization of a boundary curve of interest in $\NCreases$ dimensional fold angle space, and $\hat{\U}(s) = \U\left(\bMap(s)\right)$ be the energy along the boundary curve.
The tangent energy derivative is
\begin{equation}
    \frac{\mathrm{d} \hat{\U}}{\mathrm{d} s} = \sum_i k_i \left(\varphi_i(s) - \foldAngleO_i\right) \frac{\mathrm{d} \varphi_i}{\mathrm{d} s}.
\end{equation}
Thus, the weak variation in energy along a metastable region reflects a near cancellation between the elastic work released by some creases and that stored by others along the allowable folding motion.
The corresponding effective tangential stiffness is
\begin{equation}
    \frac{\mathrm{d}^2 \hat{\U}}{\mathrm{d} s^2} = \dphi \cdot \left(\mathbf{K} \dphi\right) + \nabla \U \cdot \ddphi,
\end{equation}
where $\mathbf{K} = \nabla^2 \U = \diag\left(k_1, \dots, k_\NCreases\right)$, and where, along metastable curves, $\mathrm{d}^2 \hat{\U} / \mathrm{d} s^2 \approx 0$.
The physical interpretation of the metastability of the boundary curve in symmetry-reduced space depends on its physical underpinning.
When the boundary corresponds to self-contact, the normal component of $-\nabla \U$ is balanced by a contact reaction.
If the associated contact multiplier is positive, feasible perturbations that open the contact increase the energy to first order, producing a sharp one-sided minimum transverse to the boundary, analogous to the multi-configuration rigidity described by Dorn, Li, and Pellegrino \cite{dorn2023multi,dorn2021structures}.
Along the boundary, however, the crease elastic work contributions nearly cancel, leaving the structure energetically soft in the tangential direction.
The resulting state is therefore pinned transverse to the boundary but soft along it.
This behavior does not necessarily correspond to a kinematic floppy mode and, when a boundary arises from the restriction to a symmetry-preserving kinematic branch only, the pinning does not by itself establish stability against symmetry-breaking perturbations.\footnote{\hl{Only $1$ dimensional metastable regions are considered herein, but higher dimensional analogs can similarly be defined by considering higher dimensional parameterizations, $\bMap\left(s_1, s_2, \dots, s_n\right)$, and considering gradients, Hessians, etc., of $\hat{\U}$ with respect to the parameters $s_1, \dots, s_n$.}} 
}

\paragraph*{\hl{Admissible space and its topology.}}
Changing $\sectAngle_1$ adjusts the way the vertex folds and, consequently, the regions in configuration space for where self-intersection occurs.
Interestingly, the topology of the admissible configuration space also goes through transitions\footnote{Similar phenomena regarding ``topological kinematics'' was also explored in Liu et al.~\cite{liu2018topological}}.
For the $\sectAngle_1 = \pi / 4$ case, admissible space consists of two separate regions connected by a point at the flat state, $\left(0, 0\right)$.
Restricted to $\left[-\pi, \pi\right] \times \left[-\pi, \pi\right]$, there are holes in the top left, top right, bottom left, and bottom right of configuration space.
The holes are exterior to the main admissible regions.
For the $\sectAngle_1 = \pi / 3$ case, there are still two main regions of admissible configuration space connected by a point at the flat state, but, notably, there are also $1$ dimensional manifolds (depicted by dashed lines) in the top left and bottom right that also make connections between the two main admissible regions.
These two $1$-handles 
are approximately given by
\begin{equation}
	\begin{split}
		\left\{-\pi\right\} &\times (\approx\pi / 2, \pi] \cup [-\pi, \approx -\pi / 2) \times \left\{\pi\right\}, \qquad \text{and} \\
		\left\{\pi\right\} &\times (\approx-\pi / 2, -\pi] \cup (\approx -\pi / 2, \pi] \times \left\{-\pi\right\},
	\end{split}
\end{equation}
respectively.
Because of the $1$-handles, two of the holes are now in the ``interior'' of admissible regions.
For the $\sectAngle_1 = 5 \pi / 12$, the topology changes further.
Here the $1$-handles appear to have expanded such that the two main admissible regions are larger and connected at three points: $\left(-\pi, \pi\right)$, $\left(\pi, -\pi\right)$, and the flat state.
These topological transitions have important implications for multistability, as seen in comparing the number of stable configurations for fixed $\rk$ but varying $\sectAngle_1 = \pi /4, \pi / 3,$ and $5 \pi / 12$.
For $\rk = 1$, for instance, the number of symmetry-constrained stable states are $3$, $8$, and $3$, respectively.
In particular, the $1$-handles for the $\sectAngle_1 = \pi / 3$ case provide an opportunity for achieving multiple stable states with energetic frustration.
The energy given in \eqref{eq:energy} is a Morse function~\cite{matsumoto2002introduction,milnor1963morse}.
The application of Morse theory to gain a deeper insight into the connections between multistability and the topology of admissible kinematics is a potentially interesting topic for future work.
\newcommand{\stabFigureWidth}{0.82\linewidth}
\begin{figure}
	\centering
	\includegraphics[width=\stabFigureWidth]{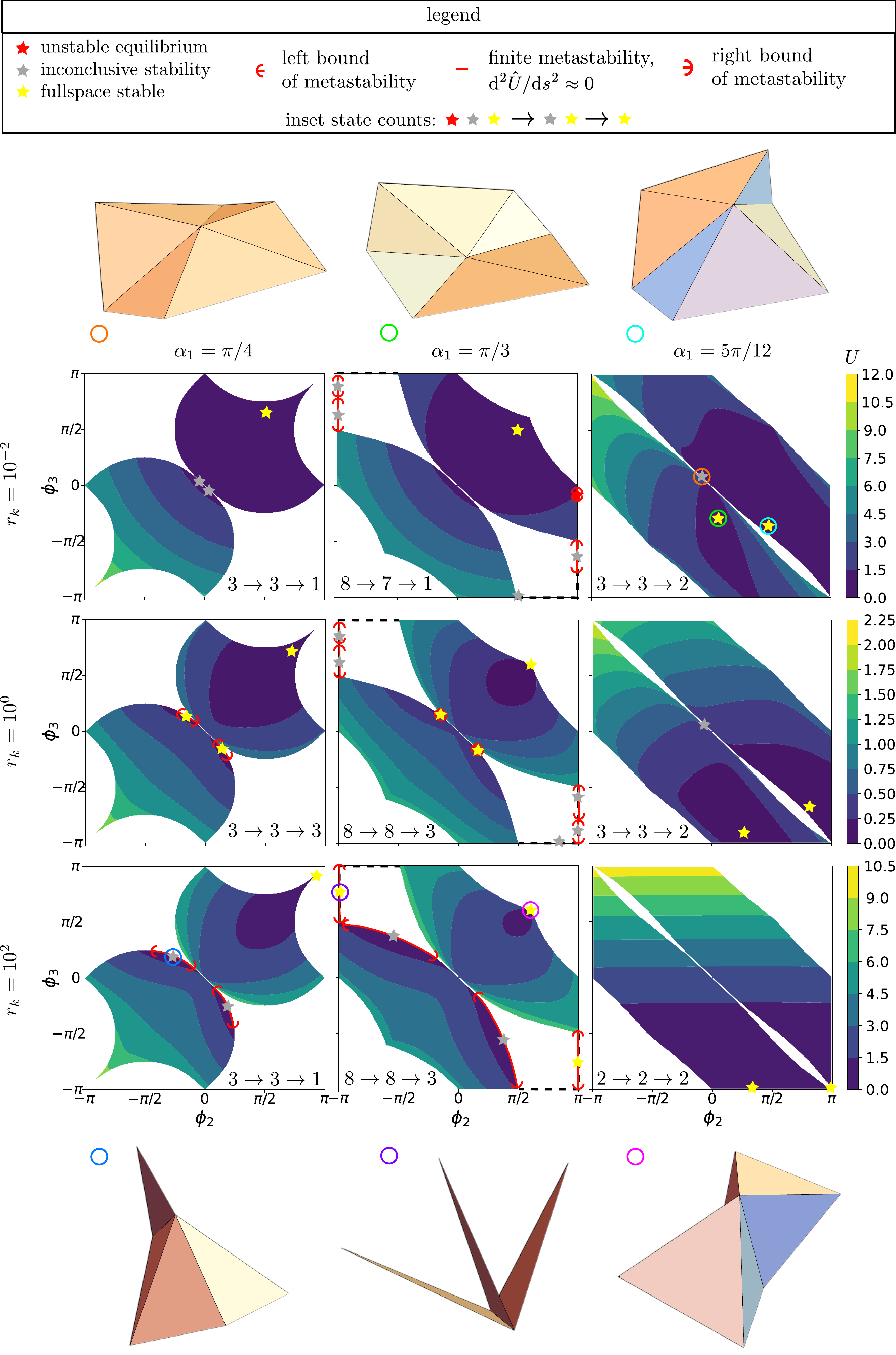}
	\caption{
		\textbf{Multistability, metastability, and topological kinematics of degree-$6$ vertices.}
		\hl{Energy landscapes for $\sectAngle_1=\pi/4$, $\pi/3$, and $5\pi/12$ (columns), and $\rk=10^{-2}$, $10^0$, and $10^2$ (rows). Stars denote local minima in the $\dihedral{1}$-symmetric configuration space: gold stars are certified as full-space minima, red stars are unstable to admissible symmetry-breaking perturbations, and gray stars remain inconclusive. White regions are excluded by self-intersection, and red intervals mark metastable boundary segments. Insets report $N_{\mathrm{sym}}\rightarrow N_{\mathrm{nr}}\rightarrow N_{\mathrm{full}}$: the numbers of symmetry-constrained minima, candidates not ruled out, and certified full-space minima, respectively. 
        }
	}
	\label{fig:multistability-6}
\end{figure}

\paragraph*{\hl{Crease stiffness ratio.}}
The energy landscape is not solely dictated by $\sectAngle_1$ (i.e., the geometry of the creases).
The interplay between the crease stiffness ratios, and the topology and shape of the admissible kinematic domains influences the emergence and characteristics of energy minima and the corresponding local energy profile. 
For instance, depending on the specific topology and shape defined by the folding constraints, variations in the crease stiffness ratio, $\rk$, can give rise to extended metastable submanifolds. 
Within these ``valley-like'' regions of the energy landscape, the energy changes minimally along certain degrees of freedom, leading to a near-flat energy profile.

Consider the case of $\sectAngle_1 = \pi / 4$ (left column).
As $\rk$ is increased, the two stable states located in the bottom-left of the configuration space exhibit a notable separation. 
Simultaneously, the region of metastability about each state expands. 
\hl{Under actuation, a contact-pinned metastable channel of this type may promote stick-slip behavior (even with respect to symmetry breaking), especially in the presence of contact friction and/or other dissipative mechanisms.
Here, physically, such behavior is a consequence} of a nearly closed fold angle at crease $4$ coexisting with a local soft mode tangential to the kinematic boundary (as shown in configuration denoted the blue circle, $\textcolor{blue}{\circ}$, in \fref{fig:multistability-6}), involving coupled changes in $\foldAngle_2$ and $\foldAngle_3$.
Recent work has suggested that metastability may have certain advantages for actuation and deployment~\cite{deshpande2024golden,zhou2025hyper}.
A similar phenomenon, albeit more pronounced, is observed for $\sectAngle_1 = \pi / 3$, where increasing $\rk$ leads to larger metastable regions and even a bifurcation between $\rk = 10^0$ and $\rk=10^{-2}$, resulting in the collision and annihilation of the two stable states. 
Conversely, for $\sectAngle_1 = 5 \pi / 12$, no metastable regions are observed across varying $\rk$. 
However, this case still exhibits interesting behavior, with one of the stable states near the flat configuration vanishing at the high stiffness ratio, $\rk = 10^2$ (orange circle, $\textcolor{myorange}{\circ}$, in \fref{fig:multistability-6}).

\hl{Reduced-dimensional} mapping of the energy landscape allowed for \begin{inparaenum}[1)] \item an exhaustive search of stable states within the slice of configuration space for the symmetry of interest, and \item the ability to probe the energy profile local to each stable state, and identify regions of metastability where the energy landscape is effectively flat tangent to the boundary of admissible space. \end{inparaenum}
Recall that stable states found in these slices of configuration space may not be stable in the context of the broader configuration space where symmetry breaking is allowed.
However, stable states in the reduced space are necessarily \emph{critical points} in the unconstrained space because the symmetry imposed on the kinematics matches the symmetry of the mechanical properties of the vertex.
The critical points may, however, be saddle points.
Notably, these points are of interest in finding minimum energy paths between stable states~\cite{zhou2023low}.
Therefore, the approach developed herein represents, in principle, \hl{a dimensionally reduced} screening method for candidate stable states (and energy barriers between stable states) in the general configuration space.
In other words, one can search for stable states with a given symmetry by first searching for them in the slice and then checking each of them with respect to the stability conditions in the general configuration space.
This is akin to dimensional reduction techniques in unsupervised machine learning to find phases of matter~\cite{wetzel2017unsupervised} or novel behaviors~\cite{fuchi2021design}.
Since symmetry often drives mechanics, symmetry slices of configuration space are a natural choice for such a dimensionally reduced search.

\subsection{Multistability of degree-$8$ vertices with reflection \& rotation symmetry ($\dihedral{2}$ symmetry)}

\newcommand{\auxJ}{\mathcal{J}}
\newcommand{\auxK}{\mathcal{K}}

To show the generality of this approach, we next consider the energy landscapes for degree-$8$ vertices with $\group = \genBy{\refl{\etwo}, \rotate{\ethree}{\pi}} \cong \dihedral{2}$ symmetry.
This means $\CreaseVec_1 = -\CreaseVec_5 = \eone$, $\CreaseVec_6 = \refl{\etwo} \CreaseVec_4$, $\CreaseVec_7 = -\CreaseVec_3 = -\etwo$, and $\CreaseVec_8 = \refl{\etwo} \CreaseVec_2$.
As before, we parameterize the geometry of the vertex by a single sector angle, $\sectAngle_1$, and then, by assumption, $\sectAngle_4 = \sectAngle_5 = \sectAngle_8 = \sectAngle_1$ and $\sectAngle_2 = \sectAngle_3 = \sectAngle_6 = \sectAngle_7 = \pi / 2 - \sectAngle_1$.
In this case, the reflection-rotation reduced-order compatibility condition, \eqref{eq:rotation-reflection-condition}, can be solved analytically as~\cite{grasinger2024lagrangian}
\begin{equation}
\foldAngle_1 = \left\{
   \begin{array}{r}
   \left\{
   \begin{array}{rrr}
    2\arctan\Bigg(& \auxK \cos\sectAngle \left(\cos\foldAngle_2-1\right) - \auxJ\left(\cos^2\sectAngle \cos\foldAngle_2 + \sin^2\sectAngle\right), \: & \\
    & \cos\sectAngle \sin\foldAngle_2 \auxJ - \auxK \left(\cos^2\sectAngle \cos\foldAngle_2 + \sin^2\sectAngle\right) \tan\left(\frac{\foldAngle_2}{2}\right) & \Bigg), \quad \foldAngle_2 > 0 \\
     -2\arctan\Bigg(& -\auxK \cos\sectAngle \left(\cos\foldAngle_2-1\right) - \auxJ\left(\cos^2\sectAngle \cos\foldAngle_2 + \sin^2\sectAngle\right), \: & \\
    & -\cos\sectAngle \sin\foldAngle_2 \auxJ - \auxK \left(\cos^2\sectAngle \cos\foldAngle_2 + \sin^2\sectAngle\right) \tan\left(\frac{\foldAngle_2}{2}\right) & \Bigg), \quad \foldAngle_2 < 0
   \end{array} 
   \right. 
   \\
   \left\{
   \begin{array}{rrr}
      -2\arctan\Bigg(& \auxK \cos\sectAngle \left(\cos\foldAngle_2-1\right) + \auxJ\left(\cos^2\sectAngle \cos\foldAngle_2 + \sin^2\sectAngle\right), \: & \\ 
       & \cos\sectAngle \sin\foldAngle_2 \auxJ + \auxK \left(\cos^2\sectAngle \cos\foldAngle_2 + \sin^2\sectAngle\right) \tan\left(\frac{\foldAngle_2}{2}\right) & \Bigg), \quad \foldAngle_2 > 0 \\
     2\arctan\Bigg(& -\auxK \cos\sectAngle \left(\cos\foldAngle_2-1\right) + \auxJ\left(\cos^2\sectAngle \cos\foldAngle_2 + \sin^2\sectAngle\right), \: & \\
      & -\cos\sectAngle \sin\foldAngle_2 \auxJ + \auxK \left(\cos^2\sectAngle \cos\foldAngle_2 + \sin^2\sectAngle\right) \tan\left(\frac{\foldAngle_2}{2}\right) & \Bigg), \quad \foldAngle_2 < 0
      \end{array} 
      \right.
   \end{array} \right.
\end{equation}
where
\begin{equation}
\begin{split}
\auxJ\left(\elevAngle, \foldAngle_2\right) &\coloneqq \sqrt{5 - 8\cos\elevAngle + 4\cos\foldAngle_2 - \cos\left(2\foldAngle_2\right) + 8\cos\left(4\sectAngle\right) \sin^4\left(\frac{\foldAngle_2}{2}\right)}, \\
\auxK\left(\elevAngle, \foldAngle_2\right) &\coloneqq 4\left|\cos\sectAngle \sin\elevAngle \sin\foldAngle_2\right|\sin\sectAngle \cot\elevAngle,
\end{split}
\end{equation}
and, consequently, the solution is only real valued (and therefore valid) provided: $\auxJ \in \Reals$.
The two inner cases represent two equally valid branches of folding motion.
The two branches are related through an ``odd symmetry''.
The second branch is equivalent to the first upon negating the elevation angle and all of the fold angles (i.e., $\elevAngle \rightarrow -\elevAngle, \foldAngle_i \rightarrow -\foldAngle_i$); this corresponds with rigidly flipping the vertex over~\cite{grasinger2024lagrangian}.
Note that, although the kinematics are oddly symmetric in this way, the energy landscape is not the same (assuming the rest fold angles are held fixed).

The mountain-valley folds are chosen to alternate, consistent with waterbomb origami~\cite{hanna2014waterbomb,hanna2015force,grasinger2022multistability,gillman2018truss}:
\begin{subequations}
\begin{align}
    \foldAngleO_1 = \foldAngleO_3 = \foldAngleO_5 = \foldAngleO_7 = -\pi / 2, \\
    \foldAngleO_2 = \foldAngleO_4 = \foldAngleO_6 = \foldAngleO_8 = \pi / 2,
\end{align}
\end{subequations}
As before, for the torsional stiffnesses, all of the creases with a mountain rest angle have stiffness $\kCrease_m = 0.025 \sqrt{\rk}$, and the ``valley'' creases have stiffness $\kCrease_v = 0.025 / \sqrt{\rk}$, where the ratio $\rk = \kCrease_m / \kCrease_v$ is varied.

\begin{figure}
	\centering
	\includegraphics[width=\stabFigureWidth]{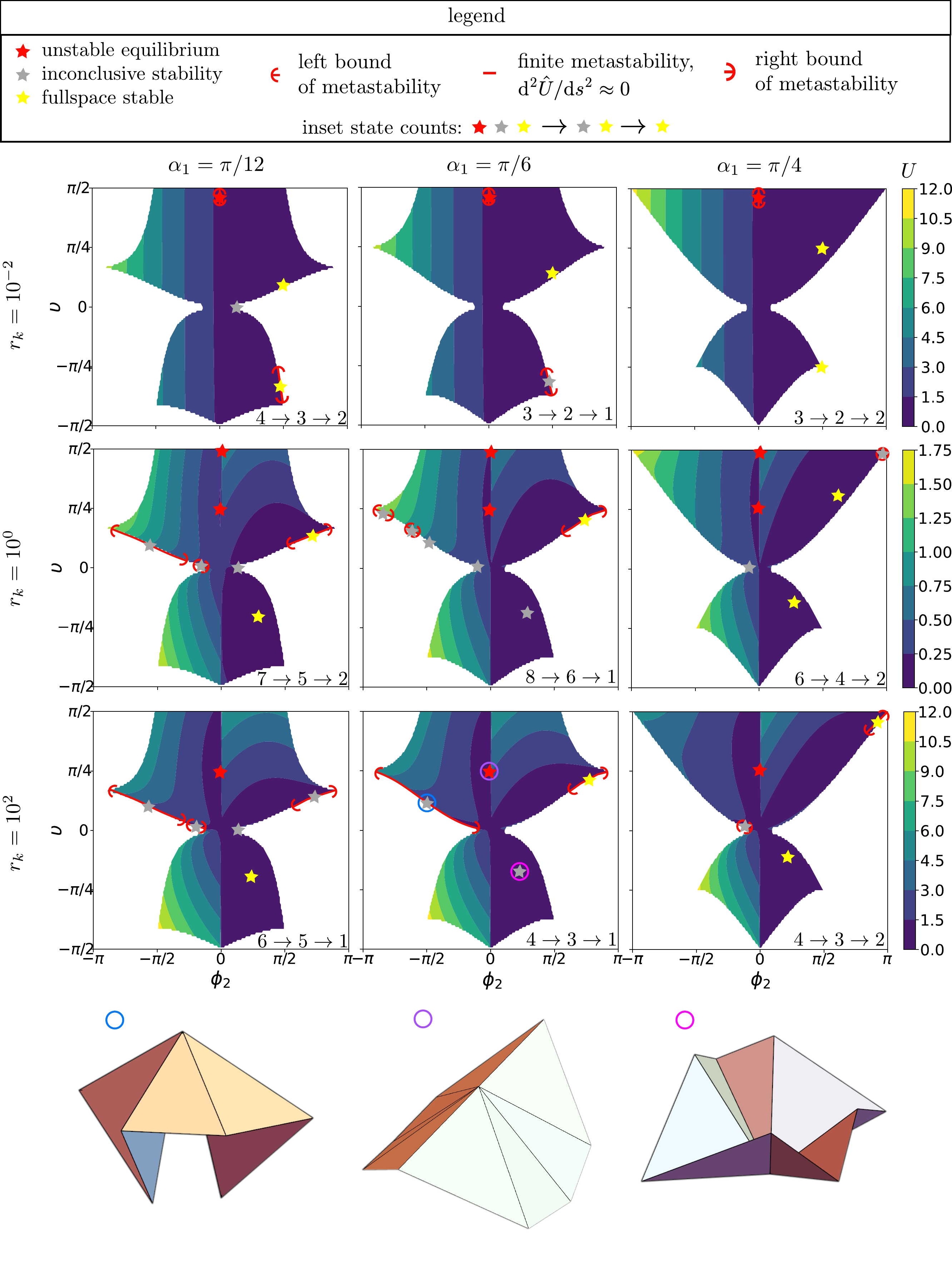}
	\caption{
		\textbf{Multistability and metastability of degree-$8$ vertices.}
		Energy landscapes for degree-$8$ vertices with $\dihedral{2}$ symmetry, with $\sectAngle_1=\pi/12$, $\pi/6$, and $\pi/4$ (columns), and $\rk=10^{-2}$, $10^0$, and $10^2$ (rows). \hl{Stars denote local minima in the symmetry-constrained configuration space: gold stars are certified as full-space minima, red stars are unstable to admissible symmetry-breaking perturbations, and gray stars remain inconclusive. White regions are excluded from the symmetry-reduced admissible domain, and red intervals mark metastable boundary segments. Insets report $N_{\mathrm{sym}}\rightarrow N_{\mathrm{nr}}\rightarrow N_{\mathrm{full}}$: the numbers of symmetry-constrained minima, candidates not ruled out, and certified full-space minima, respectively. Unlike the degree-$6$ examples, varying $\sectAngle_1$ changes the shape of the admissible boundary without inducing analogous topological transitions; shallower boundaries are associated with more extensive metastability. Colored circles identify representative configurations.}
	}
	\label{fig:multistability-8}
\end{figure}

\hl{The energy landscapes for $\alpha_1=\pi/12$, $\pi/6$, and $\pi/4$ (left, middle, and right columns, respectively) and $\rk=10^{-2}$, $10^0$, and $10^2$ (top, middle, and bottom rows, respectively) are shown in \Fref{fig:multistability-8}.
The star colors and inset triplets follow the convention introduced in \Fref{fig:multistability-6}: gold stars denote certified full-space minima, red stars denote unstable equilibria upon relaxation of the imposed symmetry, and gray stars denote cases for which the full-space classification remains inconclusive (e.g., because of contact or an inconclusive second derivative test).
Red intervals delineate metastable boundary segments.}
Here, for the $8$-vertex with $\dihedral{2}$ symmetry, the evolution of the admissible kinematic domains with changing $\sectAngle_1$ differs fundamentally from that observed in the $6$-vertex case, exhibiting no analogous topological transitions. Nonetheless, the shape of the boundary of the accessible configuration space is significantly influenced by $\sectAngle_1$. 
For instance, examining the region immediately above the flat state ($\elevAngle \gtrsim 0$), the boundary exhibits a flatter profile, characterized by a smaller slope, when $\sectAngle_1 = \pi / 12$. 
This slope progressively increases as $\sectAngle_1$ becomes larger. 
Notably, these shallower boundary slopes appear to correlate with increased metastability and a propensity for stick-slip behavior in the vicinity of the flat state, particularly for moderate and larger crease stiffness ratios.

Across all considered values of $\sectAngle_1$, the maximum number of stable states within the explored fold branch is consistently observed when the crease stiffness ratio is unity. 
It is important to note that these energy landscapes represent only one of the potential folding branches. 
While the waterbomb base is often associated with two stable configurations, our analysis reveals the possibility of a greater number of stable states arising when imposing symmetry constraints and/or within energetically frustrated regions of the domain where self-contact limits further folding. 
These additional equilibria, although potentially challenging to access experimentally, are unveiled through the low-dimensional mapping of the energy landscape, highlighting the complexity inherent in seemingly simple origami vertices.

\hl{
Each local minimum identified in the symmetry-reduced energy landscapes
was tested against all locally compatible fold-angle perturbations, including symmetry-breaking perturbations.
In aggregate, 18 of the 41 degree-6 symmetry-constrained minima and 14 of the 45 degree-8 symmetry-constrained minima remain strict minima in the full compatible space. The remaining candidates are either unstable to an admissible symmetry-breaking perturbation or remain inconclusive because of contact, a symmetry-compatibility boundary, or numerical marginality.
}

\section{Scaling to degree $\NCreases$ vertices} \label{sec:origami-cone}

\begin{figure}
	\centering
	\includegraphics[width=\linewidth]{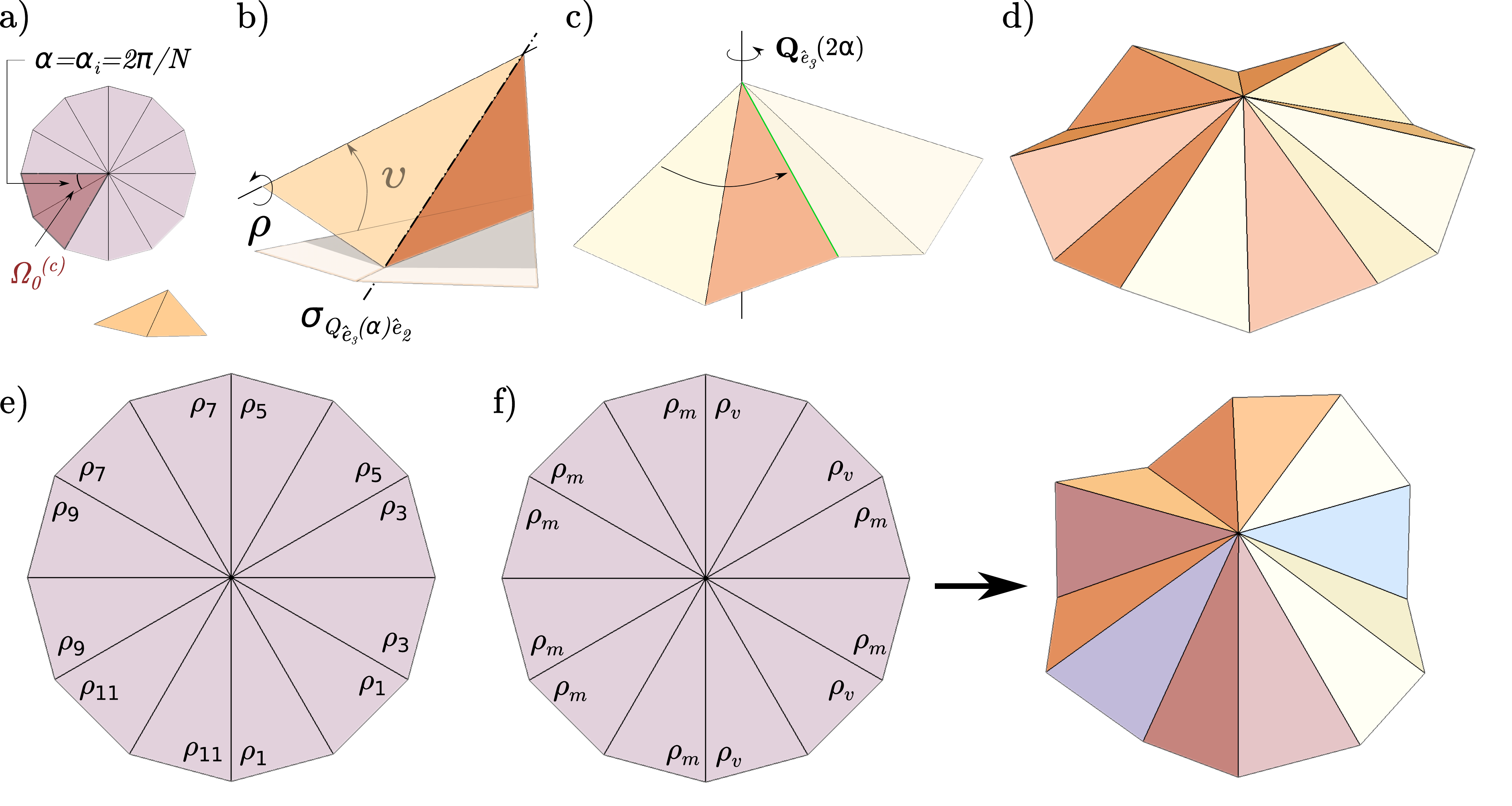}
	\caption{
		\textbf{Single degree of freedom foldings of uniform degree $\NCreases$ vertices.}
		\textbf{a)} Uniform degree $\NCreases$ vertex; unit cell $\Facet{c}$ to be folded and neighboring facet (together delimited by dashed lines) which is generated by reflecting about their shared crease; together these form a flap.
		\textbf{b)} $\Facet{c}$ is folded by elevating by angle $\elevAngle$ then rotating by $\partFoldAngle$ about its left crease; then reflecting about its right crease to generate the folded state of the flap.
		\textbf{c)} A compatibly folded vertex is then generated by taking the orbit of the flap with $\genBy{\rotate{\ethree}{2 \sectAngle}}$.
		\textbf{d)} Final folded state.
		\textbf{e)} A folded state with broken symmetry can be specified by choosing $\partFoldAngle_1, \partFoldAngle_3, \dots \partFoldAngle_{\NCreases - 1}$; again, flaps consisting of three creases are generated by reflecting about their middle crease.
		\textbf{f)} Example specifications of $\partFoldAngle_1, \partFoldAngle_3, \dots \partFoldAngle_{\NCreases - 1}$ (left) and corresponding folded vertex (right).
	}
	\label{fig:N-setup}
\end{figure}

In this section, we consider vertices with an arbitrary number of creases that are uniformly distributed such that $\sectAngle_1 = \dots = \sectAngle_\NCreases = 2 \pi / \NCreases$.
Let the unit cell to be folded be given by $\Facet{c} = \Facet{1} \cup \Facet{2}$ (see \fref{fig:N-setup}.a).
Here we introduce notation for a modified deformation map in the interest of exploring lower-dimensional kinematics with symmetry breaking.
Let $\partFoldAngle_i$ denote the angle of rotation of $\Facet{i}$ about $\CreaseVec_i$.
Then the reflection-rotation convention for the deformation map (and associated deformation gradient) takes the form
\begin{equation} \label{eq:cone-deformation}
		\F_1 = \rotate{\etwo}{\elevAngle} \rotate{\CreaseVec_1}{\partFoldAngle_1}, \quad \F_{j} = \rotate{\etwo}{\elevAngle} \rotate{\CreaseVec_1}{\partFoldAngle_1} \prod_{i=2}^{j} \rotate{\CreaseVec_{i}}{\foldAngle_{i}} \text{ for } j = 2, \dots, \NCreases.
\end{equation}
and \eqref{eq:rotation-reflection-condition} admits the solutions
\begin{equation}
	\partFoldAngle_1\left(\elevAngle\right) = \begin{cases}
		\partFoldAngle_v\left(\elevAngle\right) \\
		\partFoldAngle_m\left(\elevAngle\right)
	\end{cases},
\end{equation}
where
\begin{subequations} \label{eq:partial-N}
	\begin{align}
	\label{eq:partial-valley-N}
	\partFoldAngle_v\left(\elevAngle\right) &\coloneqq \begin{cases}
	\quad \arccos\left(\frac{\left|\cos \elevAngle + \sin \sectAngle\right|}{1 + \cos \elevAngle \sin \sectAngle}\right), & \elevAngle > 0 \\
	\quad \arccos\left(\frac{\left|\cos \elevAngle - \sin \sectAngle\right|}{1 - \cos \elevAngle \sin \sectAngle}\right), & \elevAngle < 0
	\end{cases} \\
	\label{eq:partial-mountain-N}
	\partFoldAngle_m\left(\elevAngle\right) &\coloneqq \begin{cases}
	-\arccos\left(\frac{\left|\cos \elevAngle - \sin \sectAngle\right|}{1 - \cos \elevAngle \sin \sectAngle}\right), & \elevAngle > 0 \\
	-\arccos\left(\frac{\left|\cos \elevAngle + \sin \sectAngle\right|}{1 + \cos \elevAngle \sin \sectAngle}\right), & \elevAngle < 0
	\end{cases}
	\end{align}
\end{subequations}
Given \eqref{eq:partial-N}, we can generate a compatible folded vertex by letting $\facet{1} = \F_1 \Facet{1}$, $\group = \genBy{\refl{\etwo}, \rotate{\ethree}{2 \sectAngle}}$, and $\cBody = \orbit{\facet{1}}{\group}$ (see \fref{fig:N-setup}.b-d).
One way of performing this construction which is instructive is to let $\facet{2} = \refl{\CreaseVec{2} \times \ethree} \facet{1}$ and $\facet{j} = \rotate{\ethree}{2 \pi  \left \lfloor \left(j - 1\right) / 2 \right \rfloor / \NCreases} \facet{j \mod 2}$ (\fref{fig:N-setup}.b and \fref{fig:N-setup}.c, respectively).
This is essentially the process outlined in \Fref{sec:rotation-reflection-kinematics}, and, for this case, results in a folded origami with $\dihedral{\NCreases / 2}$ symmetry.
In this case, all of the odd creases have the same fold angle, $2 \partFoldAngle_1$.
To find the fold angle at the even creases, we use the deformation map and let $\unNormalVec_1 = \creaseVec_1 \times \creaseVec_2$, $\foldAngle_2 = 2 \arccos\left(\unNormalVec_1 \cdot \left(\refl{\CreaseVec_2 \times \ethree} \unNormalVec_1\right) / \left|\unNormalVec_1\right|^2\right)$, and arrive at
\begin{equation} \label{eq:cone-phi-even}
  \foldAngle_2 = \begin{cases}
    -2 \left|\elevAngle\right|, & \foldAngle_1 = 2 \partFoldAngle_v \\
    \quad 2 \left|\elevAngle\right|, & \foldAngle_1 = 2 \partFoldAngle_m
  \end{cases};
\end{equation}
that is, the even creases have fold angle of twice the magnitude of $\elevAngle$ and opposite parity to $\foldAngle_1$.
This is a waterbomb folding for a general number of even creases.

\begin{figure}
	\centering
	\includegraphics[width=\linewidth]{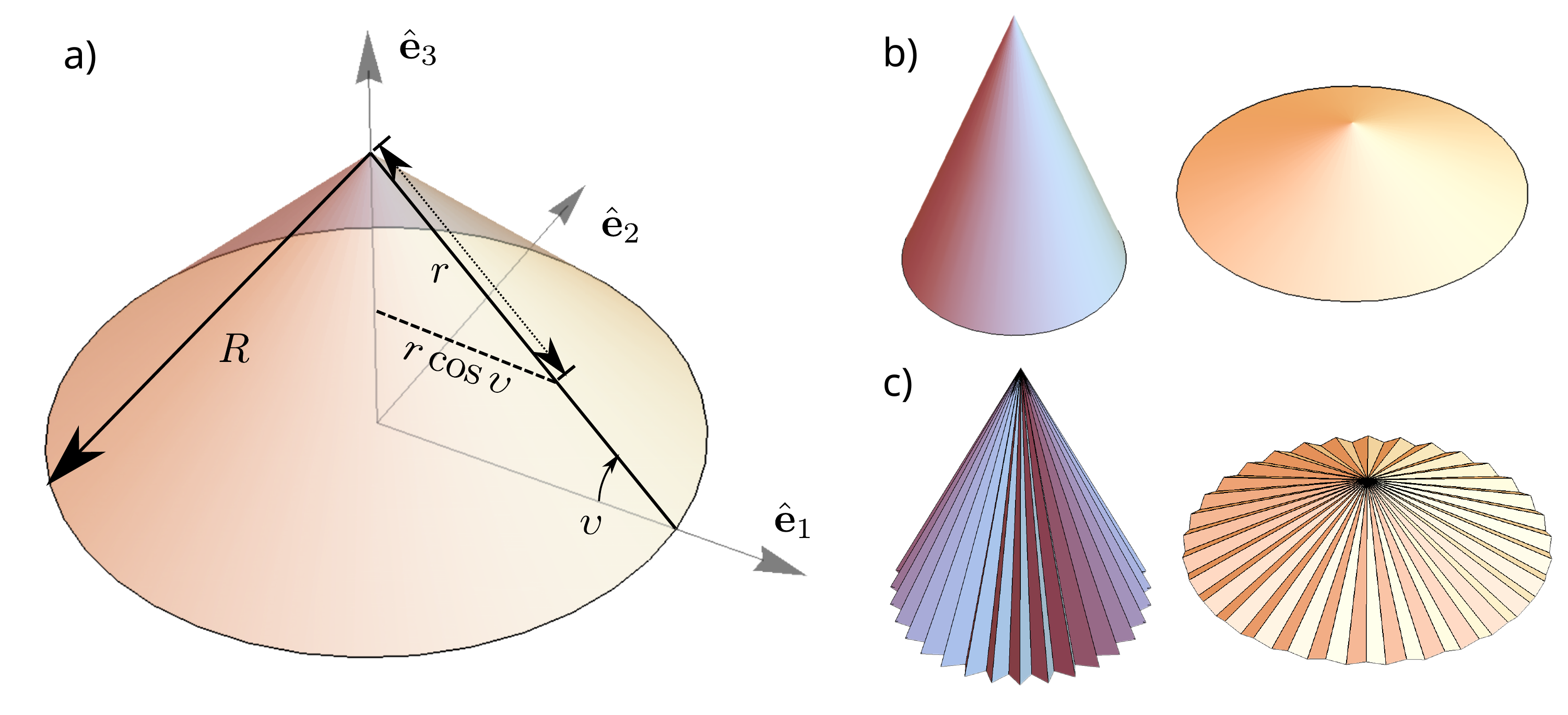}
	\caption{
		\textbf{Comparing the continuum cone to the uniform degree $\NCreases$ vertex.}
		\textbf{a)} Continuum cone and its parameterization.
		\textbf{b)} Example cone deformation when areal stretching is not penalized and axisymetry is preserved.
		\textbf{c)} Analogous origami cone. 
	}
	\label{fig:cone-analogy}
\end{figure}

\paragraph*{Infinite creases and cone inversion.}
Next we investigate the kinematics of the vertex as $\NCreases \rightarrow \infty$.
In this limit, $\sectAngle \rightarrow 0$ and $\sin \sectAngle \rightarrow 0$.
Then by \eqref{eq:partial-N}, we have that $\foldAngle_1 \rightarrow \pm 2 \elevAngle$ and $\foldAngle_2 = -\foldAngle_1$. In other words, when enforcing \eqref{eq:rotation-reflection-condition}, \emph{the kinematics of the vertex become linear in the limit of $\NCreases \rightarrow \infty$}.
The limit of $\NCreases \rightarrow \infty$ lends itself to an analogy with the inversion of a continuum cone.
It is well-known that if a thin circular sheet is supported radially and is forced at its center, that it deforms in an nonaxisymmetric way to maintain its developability; that is, symmetry breaking occurs in order to avoid the high elastic energy associated with stretching of the sheet~\cite{wang2019flexoelectricity,cerda1998conical}.
This is the so-called ``d-cone''.
However, if the sheet is free to stretch areally but each fiber from the tip of the cone to the outside radius is constrained such that its length is constant, then the cone can readily deform in the axi-symmetric manner shown in \fref{fig:cone-analogy}.b.
Let the areal stretch, which depends on $\elevAngle$, be denoted by $\arealStretch$.
As $\NCreases \rightarrow \infty$, the area that is hidden from the surface of the cone by folding about its creases, as a function of $\elevAngle$, corresponds exactly with $\arealStretch$ (see \fref{app:cone}).
This provides some intuition for why the origami cone is able to deform axisymmetrically yet remain developable.

\begin{figure}
	\centering
	\includegraphics[width=0.75\linewidth]{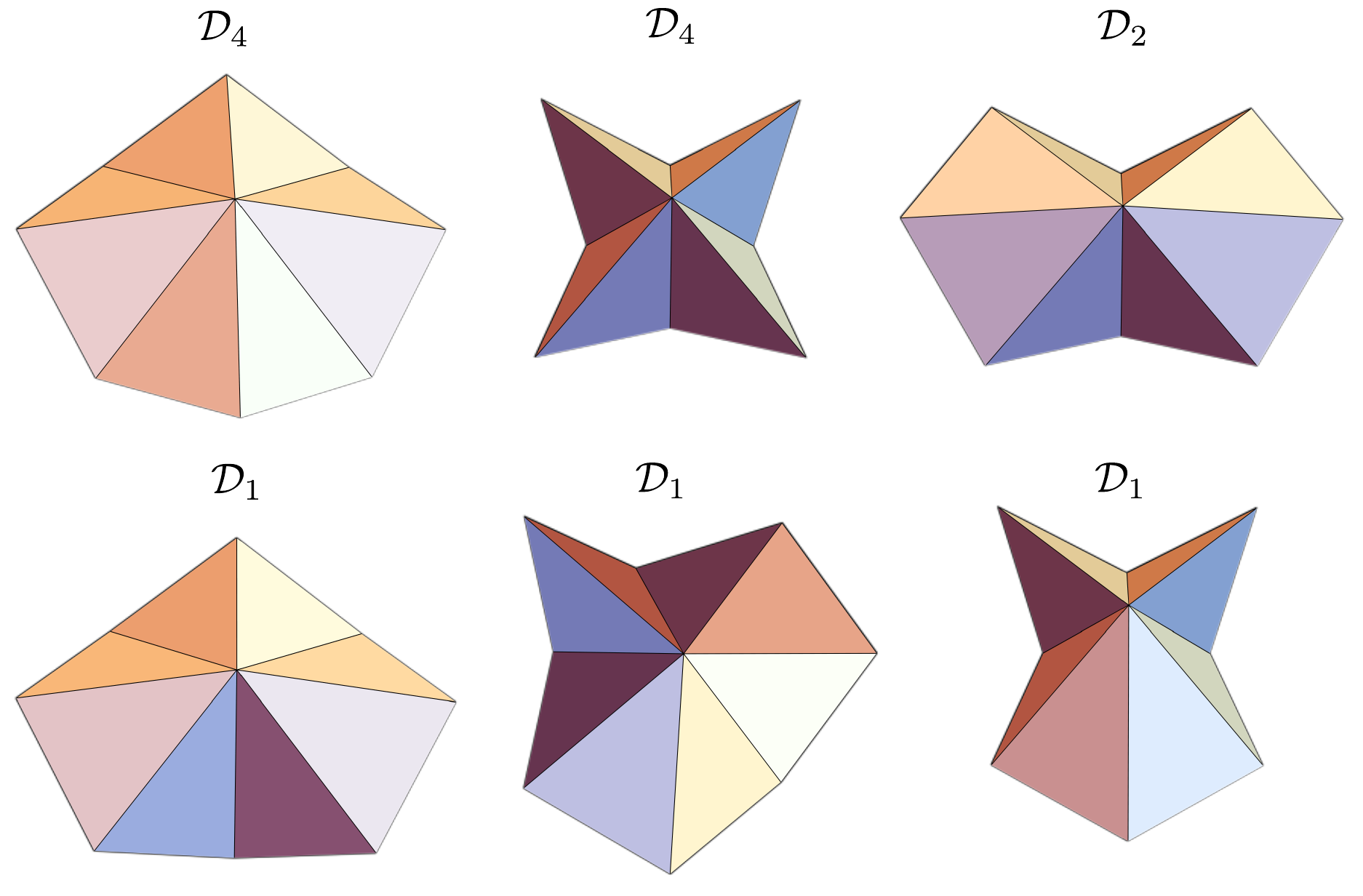}
	\caption{
		\textbf{Unique single degree of freedom solutions for the uniform degree $8$ vertex.}
		The symmetry of each solution is a subgroup of $\dihedral{4}$.
		By the orbit-stabilizer theorem, $\dihedral{4}$ solutions account for a single realization each, the $\dihedral{2}$ solution accounts for two realizations, and each of the three $\dihedral{1}$ solutions account for four realizations.
	}
	\label{fig:8-fold-symm-breaking-examples}
\end{figure}

\paragraph*{Symmetry breaking and combinatorics.}
Compatibly folded, kinematic solutions with symmetry-breaking can naturally be generated from solutions of \eqref{eq:rotation-reflection-condition}~\cite{grasinger2024lagrangian}.
This is illustrated in \fref{fig:N-setup}.
Let $\partFoldAngle_1$ be given and $\F_2 = \refl{\CreaseVec_2 \times \ethree} \F_1 \refl{\CreaseVec_2 \times \ethree}^T$; then, by \eqref{eq:cone-deformation}, $\creaseVec_3 = \rotate{\rotate{\ethree}{\sectAngle} \etwo}{\elevAngle} \CreaseVec_3$.
Thus, a compatibly folded structure with less than $\dihedral{\NCreases / 2}$ symmetry can be constructed by \begin{inparaenum}[1)] \item fixing $\elevAngle$, \item walking around the vertex counterclockwise and making a choice of $\partFoldAngle_v$ or $\partFoldAngle_m$ for the right-side of each odd crease, \item and letting $\facet{j} = \refl{\CreaseVec_j \times \ethree} \facet{j-1}$ for each $j$ even, which implies that
\begin{equation}
	\foldAngle_j = \begin{cases}
		\partFoldAngle_1 + \partFoldAngle_{\NCreases-1}, & j = 1 \\
		-2 \sgn\left(\partFoldAngle_{j-1}\right) \left|\elevAngle\right|, & j \text{ even} \\
		\partFoldAngle_{j-2} + \partFoldAngle_j, & \text{otherwise}
	\end{cases}.
\end{equation}
\end{inparaenum}
An example of this symmetry-breaking construction for a degree $12$ vertex can be seen in \fref{fig:N-setup}.e and f.
Since there are two possible choices for each $\partFoldAngle_j, \: j \text{ odd}$, for a given $\elevAngle$, this amounts to $2^{\NCreases / 2}$ different possible folded states.
However, many of them are equivalent up to a rigid body rotation.
\Fref{fig:8-fold-symm-breaking-examples} shows the six unique (up to rigid rotation) solutions and their respective symmetries for the degree $8$ vertex.
In each case, the symmetry of the solution is a subgroup of $\dihedral{4}$.
The six solutions shown in \fref{fig:8-fold-symm-breaking-examples} account for the $2^4 = 16$ possible choices of partial fold angles, $\partFoldAngle_1, \partFoldAngle_3, \dots, \partFoldAngle_{7}$, in the following way: the two $\dihedral{4}$ solutions account for $1$ choice each, the $\dihedral{2}$ solution accounts for $2$ possible choices, and each of the three $\dihedral{1}$ solutions account for $4$ choices--giving $16$ in total.
The counting is a straight forward application of the orbit stabilizer theorem.

Given that not all of the $2^{\NCreases / 2}$ represent unique solutions (up to rigid body rotations), one may wonder how many unique solutions exist for a given $\NCreases$.
This is a special case of the so-called bracelet problem in combinatorics and can be solved using Polya's enumeration theorem.
To be precise, the number of unique solutions for a degree $\NCreases$ vertex is equal to the number of bracelets with $\NCreases / 2$ beads and $2$ colors, where ``beads'' correspond to pleats in the origami cone and ``colors'' correspond to the pleat being popped up or popped down.
This total number, $\mathcal{N}$, is given by
\begin{equation}
	\mathcal{N}\left(\NCreases\right) = \begin{cases}
		\frac{1}{\NCreases} \sum_{d\left|\left(\NCreases/2\right)\right.} \psi\left(d\right) 2^{\NCreases / 2 d} + \frac{3}{4} 2^{\NCreases / 4} & \NCreases / 2 \text{ even } \\
		\frac{1}{\NCreases} \sum_{d\left|\left(\NCreases/2\right)\right.} \psi\left(d\right) 2^{\NCreases / 2 d} + \frac{1}{2} 2^{\left(\NCreases / 2 + 1\right) / 2} & \NCreases / 2 \text{ odd }
	\end{cases}
\end{equation}
where $\psi$ is Euler's totient function and by the notation $d|\left(\NCreases/2\right)$ we mean the summation is over all of the divisors, $d$, of $\NCreases / 2$~\cite{gilbert1961symmetry,riordan2014introduction}.
The beginning of the sequence is tabulated in \fref{tab:necklaces}.
\begin{table}[hbt!]
    \centering
    \bgroup
    \def\arraystretch{1.5}
    \setlength{\tabcolsep}{0.5em}
    \begin{tabular}{r | *{13}{c}}
        $\NCreases$ & 6 & 8 & 10 & 12 & 14 & 16 & 18 & 20 & 22 & 24 & 26 & 28 & 30\\
        \hline
        $\mathcal{N}$ & 4 & 6 & 8 & 14 & 20 & 36 & 60 & 108 & 188 & 352 & 632 & 1182 & 2192
    \end{tabular}
    \egroup
    \caption{
    	\textbf{Combinatorics of uniform, degree $\NCreases$ vertices.}
    	The number of unique single degree of freedom foldings, $\mathcal{N}$, for uniform vertices with $\NCreases$.}
    \label{tab:necklaces}
\end{table}

\section{Combinatoric \hl{branchwise} stability of the ``origami cone''} \label{sec:cone-multistability}

We next investigate the multistability of vertices with uniformly distributed $\NCreases$, even, number of creases.
\Fref{sec:origami-cone} developed $1$ degree of freedom kinematics for $2^{\NCreases / 2 + 1}$ folding branches.
\hl{Throughout this section, stability is understood branchwise: a configuration is counted when it is a local minimum of the crease energy with respect to motion along its prescribed one-dimensional folding branch. Such a minimum is a candidate full-space stable state but need not remain stable to compatible perturbations transverse to that branch.}
Using the torsional crease elastic energy given by \eqref{eq:energy}, we ask ``how many of the folding branches have a \hl{branchwise minimum}?''.
Recall that the folding branches generated herein exhibit a different behavior for the odd creases than the even creases (\eqref{eq:partial-N} and \eqref{eq:cone-phi-even}, respectively).
As a result, we consider mechanical properties where all of the odd creases have the same stiffness and all of the even creases have the same stiffness.
The vertices are parameterized by the ratio $\rk = \kCrease_{\text{even}} / \kCrease_{\text{odd}}$
and the angle of elevation of the stress free configuration, $\elevAngle_0$, where all of the odd folds are valley and all of the even folds are mountain (i.e., the waterbomb folding),
\begin{equation}
	\foldAngleO_i = \begin{cases}
		2 \arccos\left(\frac{\left|\cos \elevAngle_0 + \sin \sectAngle\right|}{1 + \cos \elevAngle_0 \sin \sectAngle}\right), & i \text { odd } \\
		-2 \elevAngle_0, & i \text { even }
	\end{cases}
\end{equation}
where $\sectAngle$ is the (uniform) sector angle between each pair of neighboring creases.

\begin{figure}
	\centering
	\includegraphics[width=\linewidth]{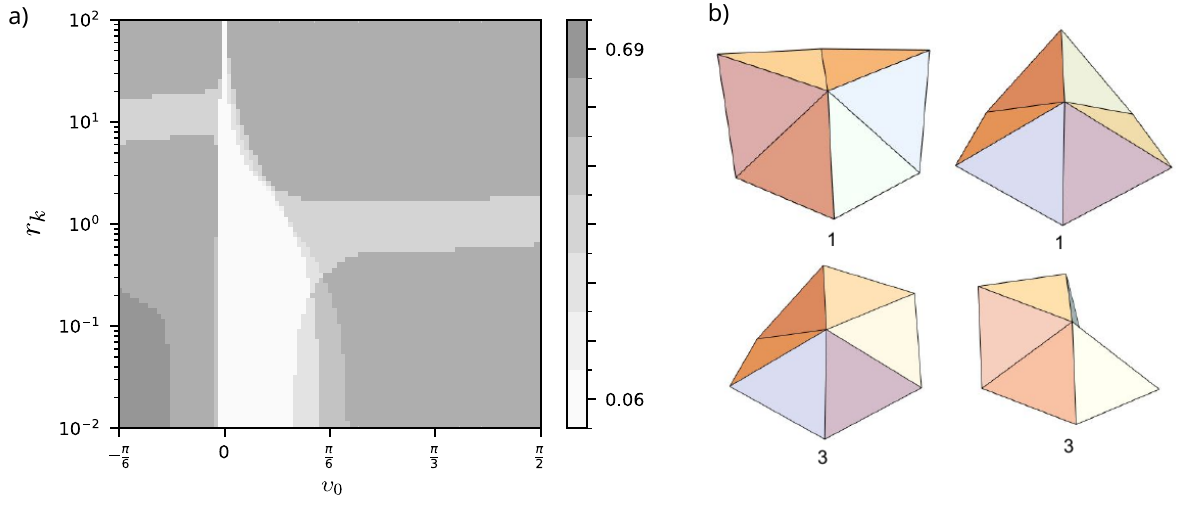}
	\caption{
		\textbf{Multistability of the $6$ fold cone.}
		\textbf{a)} Normalized number of \hl{branchwise} stable states on the $16$ folding branches as a function of $\rk$ and $\elevAngle_0$.
		The normalized number of states varies from $0.0625$ -- $0.6875$ with non-monotonic, nontrivial dependencies.
		Near $\elevAngle_0 \approx -\pi / 6$ and $\rk < 1$, there is structure with $0.25 \: (=4/16)$ and $0.4375 \: (=7/16)$ normalized number of states.
		There are bands of $0.3125 \: (=5/16)$ states at $\elevAngle_0 \lessapprox -\pi / 6$, $\rk \approx 1$, and $\elevAngle_0 > 0$, $\rk \approx 10$.
		A region with $0.6875 \: (=11/16)$ stable states is found at $\elevAngle_0 \approx \pi /6$ and $\rk \lessapprox 0.1$.
		\textbf{b)} Example snapshots of folding branches and the corresponding degeneracy of each snapshot (i.e., number branches with equivalent energy).
	}
	\label{fig:cone-3}
\end{figure}

\Fref{fig:cone-3} shows the number of \hl{branchwise} stable states for the $6$ fold ``cone'' mapped out in mechanical property space as a function of $\rk$ and $\elevAngle_0$.
The number of stable states is normalized by the number of folding branches where, for the $6$ fold cone, there are $16$ folding branches off of the flat state.
This comes from $6 / 2 = 3$ flaps that can either be folded ``up'' or ``down'', and, for each choice of flap folding, a folding branch above and below the flat state, for a total of $2^3 \times 2 = 16$.
No branch was found with more than one stable state; and the number of folding branches with a stable state varied from $0.0625$ to $0.6875$.
The stable states are found by a multi-start, line search optimization on each separate folding branch.\footnote{States are counted as \hl{branchwise} stable if $\left|\nabla \U\right| < 10^{-6}$ and the well depth, $\Delta \U$, relative to the energy of the flat state is above a threshold (i.e., $\Delta \U / \U_{\mathrm{flat}} > 10^{-3}$.)}
The multistability mapping of parameter space is restricted to $\elevAngle_0 \in \left[-\pi / 2, \pi / 6\right]$ because beyond these bounds, self-contact occurs.
The normalized number of stable states is $0.0625 \: (= 1 / 16)$ when $\elevAngle_0 = 0$, regardless of $\rk$, because the $\left|\foldAngle\right|$ is monotonically increasing with $\left|\elevAngle\right|$ for all of the constructed solutions. 
Interestingly, however, a lot of the variability in the number of stable states is found below but near to the flat state, in the $-\pi / 5 < \elevAngle_0 < -\pi / 12$ range, and when $\rk < 10$.
Here there are regions with $0.25 \: (=4/16)$ and $0.4375 \: (=7/16)$ stable states.
Near $\rk \approx 1$, there is a band denoting $0.3125 \: (=5/16)$ stable states, nearly irrespective of $\elevAngle_0$ provided $\elevAngle_0 \lessapprox -\pi / 6$.
There is a similar band near $\rk \approx 10$ for $\elevAngle > 0$.
Lastly, note the region with $0.6875 \: (=11/16)$ stable states for $\elevAngle \approx \pi /6$ and $\rk \lessapprox 0.1$.
Otherwise, a large portion of parameter space has $0.125 \: (=2/16)$ stable states.

\begin{figure}
	\centering
	\includegraphics[width=0.70\linewidth]{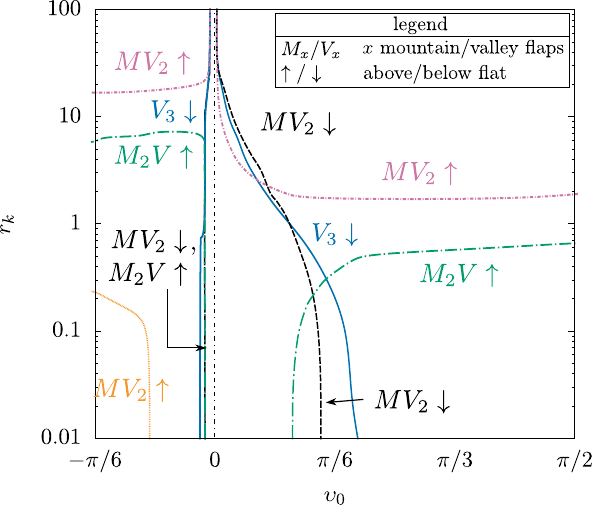}
	\caption{
		\textbf{Branch stability boundaries of the $6$ fold cone.}
		Each line represents a boundary in the parameter space defined by a stress free elevation angle ($\elevAngle_0$) and relative crease stiffness ($\rk$). Crossing one of these boundaries signifies a change in the stability of its associated folding branch. Branches are identified by labels with the following convention: $M$ indicates a partial fold angle assigned as mountain ($\partFoldAngle = \partFoldAngle_m$), and V indicates a valley assignment ($\partFoldAngle = \partFoldAngle_v$). Subscripts denote the number of such folds. 
        The symbol $\uparrow$ ($\downarrow$) signifies that the branch is above (below) the flat state (i.e., $\elevAngle > 0$ ($\elevAngle < 0$)).
        Labels are positioned on the side of the boundary where the corresponding branch is stable; generally, this is away from the $\elevAngle_0 = 0$ axis where branches tend to be less stable. Many boundaries exhibit segments that are nearly horizontal (indicating an $\rk$ threshold for stability) and/or nearly vertical (an $\elevAngle_0$ threshold). 
        The intricate pattern of these intersecting stability boundaries gives rise to the complex ``phase behavior'' and multistability regions in parameter space.
	}
	\label{fig:cone-3-firing}
\end{figure}

To further explain the various numbers of stable states in parameter space, \fref{fig:cone-3-firing} shows the stability boundaries for each of the folding branches; that is, it shows the boundaries that, when crossed, the stability of its associated folding branch changes.
These correspond to ``phase boundaries'' in \fref{fig:cone-3}.
The boundaries are labeled with an abbreviation of each branch.
$M$ signifies an assignment of $\partFoldAngle = \partFoldAngle_m$ and $V$ signifies an assignment of $\partFoldAngle = \partFoldAngle_v$.
The symbol $\uparrow$ denotes that the branch is above the flat state (i.e., $\elevAngle > 0$), and $\downarrow$ denotes below the flat state (i.e., $\elevAngle < 0$).
For instance, by $M_2 V \uparrow$ we mean that $2$ of the $\partFoldAngle$ are mountain, $1$ is valley, and the branch is above the flat state.
The boundaries are labeled from the inside; that is, the branch is stable on the side of the boundary where the label appears.
Notice that all of the labels are on the sides of the boundaries opposite to $\elevAngle_0 = 0$ because all branches tend to be less stable in this limit.
Some of the boundaries, such as $M V_2 \downarrow$ and $V_3 \downarrow$ are primarily vertical with some regions of finite slope.
However, for many of the boundaries, part of the contour is nearly horizontal -- representing an $\rk$ threshold -- and part of the contour is nearly vertical -- representing an $\elevAngle_0$ threshold.
For example, the branch $M_2 V \uparrow$ is stable for nearly every parameter set that satisfies the simple conditions $\rk < 0.7$ and $\elevAngle_0 > \pi / 8$, or $\rk < 7$ and $\elevAngle_0 < -\pi / 36$.
It is the intersecting of these many boundaries that give rise to the ``phase behavior'' observed in \fref{fig:cone-3}.

\begin{figure}
	\centering
	\includegraphics[width=\linewidth]{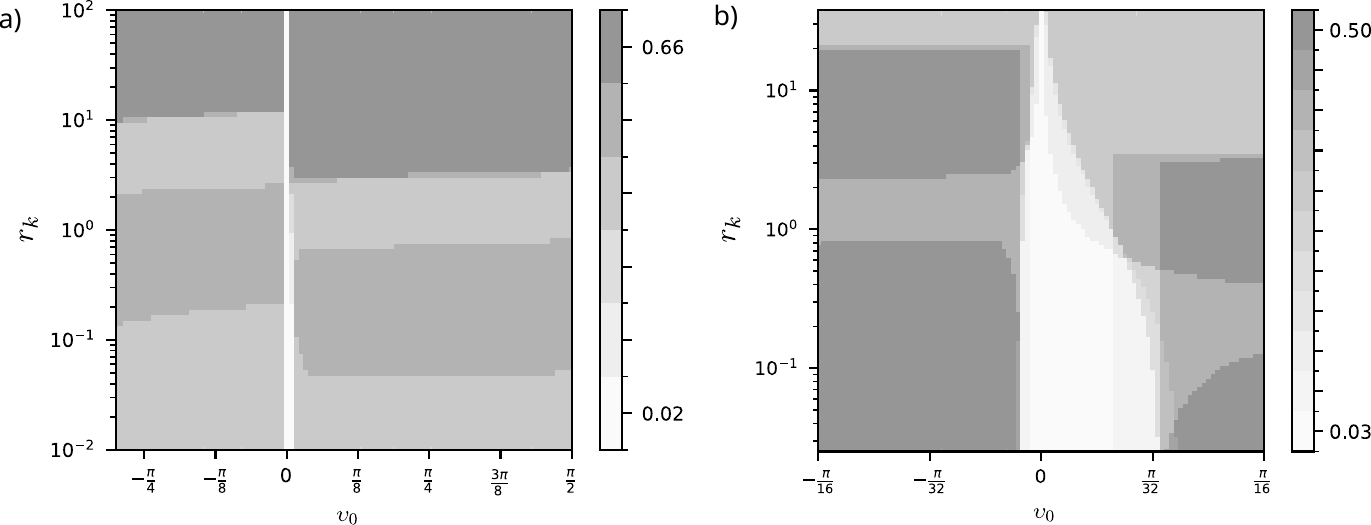}
	\caption{
		\textbf{Multistability of the $8$ fold cone.}
		\textbf{a)} Normalized number of \hl{branchwise} stable states on the $32$ folding branches as a function of $\rk$ and $\elevAngle_0$.
		The normalized number of stable states varies from $0.03125$ -- $0.5$ with non-monotonic, nontrivial dependencies.
		There are two bands each of $0.375 \: (=12/32)$ and $0.5$ normalized states for both $\elevAngle_0 < 0$ and $\elevAngle_0 > 0$.
		The number of stable states appears to be constant at $0.3125 \: (=10/32)$ above a threshold of $\rk$, provided $\elevAngle_0 \neq 0$.
		\textbf{b)} Zoomed in parameter space where a finer structure emerges.
	}
	\label{fig:cone-4}
\end{figure}

\Fref{fig:cone-4} shows the normalized number of stable states for the $8$ fold cone, which has $2^4 \times 2 = 32$ folding branches off of the flat state.
Here, the cone vertex can be folded further upward than the $6$ fold case, so $\elevAngle_0 \in \left[-\pi / 2, \pi / 4\right]$.
The normalized number of stable states varies from $0.03125 \: (=1/32)$ to $0.5$, and, similar to the $6$ fold case: \begin{inparaenum}[1)] \item there are horizontal bands (i.e., constant $\rk$) of nearly constant number of stable states, except for \item near, but below, the flat state where a finer structure with various numbers of stable states emerges. \end{inparaenum}
In contrast, there are more types of, and more total horizontal bands.
For example, there are bands of $0.375 \: (=12/32)$ and $0.5$ normalized number of states for both $\elevAngle_0$ above the flat state and below it; and the number of stable states appears to be constant at $0.3125 \: (=10/32)$ above a threshold of $\rk$, provided $\elevAngle_0 \neq 0$.
The finer structure of various stable states for the $8$ fold cone can be seen more clearly in \fref{fig:cone-4}.b where we have zoomed in to parameter space about the stress free configuration near the flat state (i.e., $-\pi / 16 \leq \elevAngle_0 \leq \pi / 16$) and with crease stiffness ratio nearer to unity (i.e., $0.05 \leq \rk \leq 25$).
There is more structure--or more options for number of stable states--than in the $6$ fold case, but it occurs over a narrower interval of $\elevAngle_0$.

\begin{figure}
	\centering
	\includegraphics[width=0.70\linewidth]{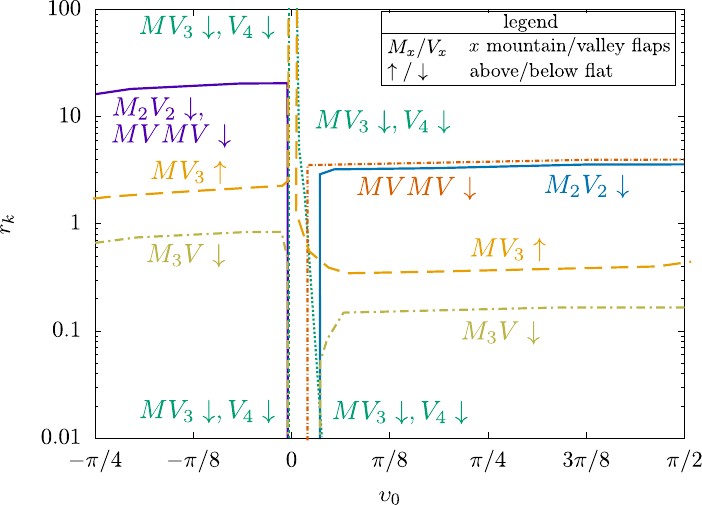}
	\caption{
		\textbf{Branch stability boundaries of the $8$ fold cone.}
		Each line delineates a boundary where the stability of an associated folding branch changes within the parameter space defined by stress free elevation angle ($\elevAngle_0$) and relative crease stiffness ($\rk$). Branch labels indicate specific fold assignments. 
        Notably, the pattern of assignments is significant: branches with the same total number of mountain and valley folds, such as $M_2 V_2$ and $MVMV$ are distinct and exhibit different stability regions.
        The stability boundaries prominently feature nearly horizontal segments, representing $\rk$ thresholds, and nearly vertical segments, concentrated near the $\elevAngle_0 = 0$ axis, representing $\elevAngle_0$ thresholds. This characteristic is more pronounced for this $8$-fold cone compared to the $6$-fold cone case.
	}
	\label{fig:cone-4-firing}
\end{figure}

\Fref{fig:cone-4-firing} shows the stability boundaries for each of the folding branches of the $8$ fold cone.
Note that, despite having the same number of mountain / valley partial fold assignments, $M_2 V_2$ and $M V M V$ are distinct branches with differing fold angles.
Again, and more prominently than in the $6$ fold cone case, part of each boundary is nearly horizontal -- representing an $\rk$ threshold -- and part of the contour is nearly vertical -- representing an $\elevAngle_0$ threshold.
The vertical portions of the boundaries are more concentrated near $\elevAngle_0 = 0$.

\begin{figure}
	\centering
	\includegraphics[width=\linewidth]{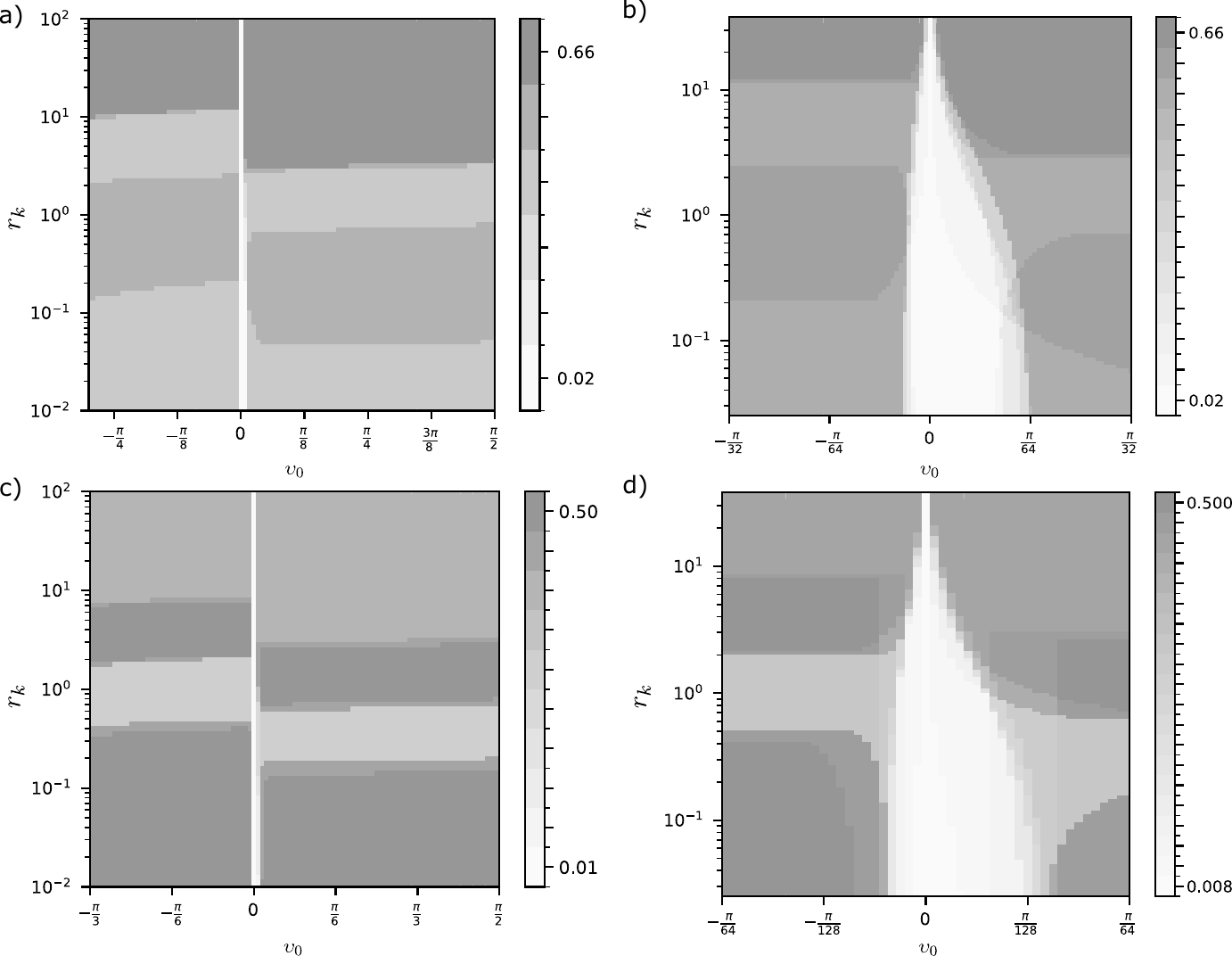}
	\caption{
		\textbf{Multistability of the $10$ and $12$ fold cones.}
		\textbf{a)} Normalized number of \hl{branchwise} stable states on the $64$ folding branches.
		The number of stable states varies from $0.015625$ -- $0.65625$ (i.e., $1/64$ -- $42/64$) with non-monotonic, nontrivial dependencies.
		There are two bands each of $0.265625 \: (=17/64)$ and $0.5$ normalized number of states for both $\elevAngle_0 < 0$ and $\elevAngle_0 > 0$.
		The normalized number of states appears to be constant at $0.65625 \: (=42/64)$ above a threshold of $\rk$, provided $\elevAngle_0 \neq 0$.
		\textbf{b)} Zoomed in parameter space where a finer structure emerges.
		\textbf{c)} Number of \hl{branchwise} stable states on the $128$ folding branches.
		The number of stable states varies from $0.0078125$ -- $0.5$ (i.e, $1/128$ -- $64/128$) with non-monotonic, nontrivial dependencies.
		There are many horizontal bands consisting of $0.336 \: (\approx 43/128)$, $0.453 \: (\approx 58/128)$, and $0.5$ normalized number of states.
		The number of stable states appears to be constant at $0.4375 \: (=56/128)$ above a threshold of $\rk$, provided $\elevAngle_0 \neq 0$.
		\textbf{d)} Zoomed in parameter space where a finer structure emerges.
	}
	\label{fig:cone-56}
\end{figure}

\Fref{fig:cone-56} shows the normalized number of stable states for the $10$ and $12$ fold cones, which have $2^6 = 64$ and $2^7 = 128$ folding branches, respectively.
Facet contact occurs at $\elevAngle = 3 \pi / 10$ and $\pi / 3$, respectively, and a general expression for when facet contact occurs is $\elevAngle = \pi / 2 \left(1 - 4 / \NCreases\right)$, which is obtained by solving $\partFoldAngle_v = \pi / 2$ using \eqref{eq:partial-valley-N}.
The trends continue where more horizontal bands appear with increasing $\NCreases$, and a finer, but narrower (in $\elevAngle_0$ space) structure emerges just below $\elevAngle_0$ at the flat state.
As supported by figures \ref{fig:cone-3-firing} and \ref{fig:cone-4-firing}, the finer structure for each of the $6$, $8$, $10$, and $12$ fold cases is due to the intersecting of continuous boundaries of stability for each of the folding branches cutting through parameter space.
Because of this, the regions in \fref{fig:cone-3}--\ref{fig:cone-56} may be considered analogous to ``phases'' and may have interesting connections to the celebrated Gibbs phase rule~\cite{sun2021generalized}.
This poses a potentially interesting topic to explore in future work.

The results presented herein can be considered both a \hl{dimensionally reduced} screening method for candidate stable states in the general configuration space, and a look at how one may achieve a number of stable states if certain constraints related to the symmetry can be enforced.

\section{Conclusion} \label{sec:conclusion}
In this work, we utilized the Lagrangian approach to origami, combined with symmetry principles, to effectively reduce the complexity and dimensionality of origami vertex kinematics. By simplifying the folding space, to $2$ dimensions or less, we were able to visualize the energy landscape. 
\hl{
This visualization enabled an exhaustive mapping of symmetry-constrained minima and metastable regions, followed by a full-space stability analysis of the degree-$6$ and degree-$8$ candidates.
}

Our analysis revealed that for degree-$6$ vertices, topological changes in the allowable kinematic domain lead to distinct and varying multistability properties. Furthermore, for both the degree-$6$ and degree-$8$ vertices investigated, these identified metastable regions may give rise to stick-slip behavior, an outcome influenced by the interplay between kinematic boundaries and crease mechanics. We also utilized a method for generating lower-dimensional kinematics that incorporates symmetry breaking to explore the rich \hl{combinatorics and mechanical property dependence of branchwise energetic minima in origami cones}.

The exact analytical solutions derived for these systems are expected not only to deepen our understanding of origami mechanics but also to facilitate and inspire new design algorithms. The diverse multistability properties uncovered have significant implications for a range of applications, including deployable structures, mechanical metamaterials, mechanical computing, origami-based robotics, mechanical energy absorption, and structures designed to self-deploy and retain their shape.
\hl{Experimental realization and validation of selected vertices predicted to exhibit multiple stable states and contact-pinned metastable regions--including the effects of finite facet compliance and thickness, contact friction, and manufacturing tolerances--present an interesting direction for future work.
More broadly, while the findings presented herein were focused on single vertices with an even number of creases, the developed approach holds considerable promise for generalization to multi-vertex crease patterns and odd-degree vertices.}

\begin{hlbreakable}
\section*{Use of artificial intelligence}
ChatGPT (GPT-5.6 Sol) was used during revision of this manuscript as an editorial and programming aid. It assisted with manuscript organization and language editing for clarity, concision, and precision of the mathematical exposition; helped refine the formulation and presentation of the full-space stability analysis; and generated initial drafts of the associated Python code and Appendix~\ref{app:full-space-stability}. All AI-assisted text, mathematics, and code were critically reviewed, modified, tested, and validated by the authors. The authors performed the final analyses and interpretation, made all scientific judgments, and take full responsibility for the content of the article.

\section*{Data accessibility}
Code and data supporting this article are archived in Zenodo at
\url{https://doi.org/10.5281/zenodo.22719444}.
Numerical data underlying the figures can be regenerated from the deposited source code and parameter files: Mathematica notebooks and python code map out the symmetry-reduced energy landscapes and perform full-space stability analysis, respectively.
The corresponding development repository is available at
\url{https://github.com/grasingerm/RSPA-2026-0429}.
\end{hlbreakable}

\section*{Acknowledgements}
We are grateful to Richard James and Kaushik Bhattacharya for many insightful discussions.
We acknowledge the support of the Laboratory University Collaboration Initiative sponsored by the US Office of the Secretary of Defense and the support of the Air Force Research Laboratory.

\appendix
\numberwithin{equation}{section}
\renewcommand{\theequation}{\thesection\hskip2pt\arabic{equation}}

\hl{
\section{Full-space stability of symmetry-reduced candidates}
\label{app:full-space-stability}

The symmetry-reduced energy landscapes in Section~3 provide candidate equilibria within prescribed symmetry classes. To determine whether these candidates remain stable when symmetry-breaking perturbations are permitted, we test each candidate locally on the full rigid-folding compatibility manifold. The procedure follows the constraint-projection approach introduced by Tachi~\cite{tachi2009simulation}, but is used here to evaluate the local energy gradient and Hessian rather than to generate a folding trajectory.

Let

\begin{equation}
    \boldsymbol{\varphi}^{\star}
    =
    \left(
        \varphi_1^{\star},\ldots,\varphi_{\NCreases}^{\star}
    \right)
\end{equation}

denote a candidate state satisfying the loop-closure condition

\begin{equation}
    \mathbf F_{\NCreases}
    \left(\boldsymbol{\varphi}^{\star}\right)
    =
    \prod_{i=1}^{\NCreases}
    \mathbf Q_{\hat{\mathbf b}_i}
    \left(\varphi_i^{\star}\right)
    =
    \mathbf I.
\end{equation}

Using the notation of Section~2, define the deformed crease directions

\begin{equation}
    \hat{\mathbf c}_i^{\star}
    =
    \mathbf F_{i-1}
    \left(\boldsymbol{\varphi}^{\star}\right)
    \hat{\mathbf b}_i
\end{equation}

and assemble them as the columns of

\begin{equation}
    \mathbf C^{\star}
    =
    \begin{bmatrix}
        \hat{\mathbf c}_1^{\star} &
        \cdots &
        \hat{\mathbf c}_{\NCreases}^{\star}
    \end{bmatrix}
    \in \mathbb R^{3\times\NCreases}.
\end{equation}

The first variation of loop closure satisfies
\begin{equation}
\delta\mathbf F_{\NCreases}
\mathbf F_{\NCreases}^{T}
=
\left[
\mathbf C^{\star}\delta\boldsymbol{\varphi}
\right]_{\times},
\label{eq:linearized-fullspace-closure}
\end{equation}
where \([\mathbf a]_{\times}\mathbf v=\mathbf a\times\mathbf v\). Consequently, the infinitesimally compatible fold-angle perturbations satisfy
\begin{equation}
\mathbf C^{\star}\delta\boldsymbol{\varphi}
=
\mathbf 0.
\label{eq:fullspace-tangent}
\end{equation}
At a regular single-vertex configuration,
\(\operatorname{rank}\mathbf C^{\star}=3\), and the tangent space therefore has dimension \(\NCreases-3\).

A vector in \(\ker\mathbf C^{\star}\) satisfies compatibility only to first order. Finite perturbations are therefore projected back onto the nonlinear loop-closure manifold. Let the columns of

$$
    \mathbf Z\in
    \mathbb R^{\NCreases\times(\NCreases-3)}
    \qquad\text{and}\qquad
    \mathbf Y\in
    \mathbb R^{\NCreases\times3}
$$

be orthonormal bases for
\(\ker\mathbf C^{\star}\) and
\(\operatorname{range}[(\mathbf C^{\star})^{T}]\), respectively, as obtained from a singular-value decomposition. A nearby fold-angle state is written as
\begin{equation}
\boldsymbol{\varphi}(\mathbf q,\boldsymbol{\eta})
=
\boldsymbol{\varphi}^{\star}
+
\mathbf Z\mathbf q
+
\mathbf Y\boldsymbol{\eta},
\label{eq:fullspace-local-coordinates}
\end{equation}
where \(\mathbf q\in\mathbb R^{\NCreases-3}\) parameterizes tangent perturbations. For each sufficiently small \(\mathbf q\), the normal correction
\(\boldsymbol{\eta}=\boldsymbol{\eta}(\mathbf q)\) is obtained by solving
\begin{equation}
\mathbf r\left(
\boldsymbol{\varphi}
\left(\mathbf q,\boldsymbol{\eta}\right)
\right)
=
\mathbf 0,
\qquad
\mathbf r(\boldsymbol{\varphi})
:=
\operatorname{ax}
\left\{
\log
\mathbf F_{\NCreases}(\boldsymbol{\varphi})
\right\},
\label{eq:fullspace-retraction}
\end{equation}
where \(\operatorname{ax}\) maps a skew-symmetric matrix to its axial vector. Equation~\eqref{eq:fullspace-retraction} defines a local retraction onto exact loop closure and gives the compatibility-restricted energy
\begin{equation}
\widetilde{\U}(\mathbf q)
=
\U\left(
\boldsymbol{\varphi}
\left(
\mathbf q,\boldsymbol{\eta}(\mathbf q)
\right)
\right).
\label{eq:fullspace-reduced-energy}
\end{equation}

The gradient \(\mathbf g\) and Hessian \(\mathbf H\) of
\(\widetilde{\U}\) at \(\mathbf q=\mathbf 0\) are evaluated using centered finite differences. For the orthonormal coordinate vectors \(\mathbf e_a\),
\begin{align}
g_a
&\simeq
\frac{
\widetilde{\U}(h\mathbf e_a)
-
\widetilde{\U}(-h\mathbf e_a)
}{2h},
\\
H_{aa}
&\simeq
\frac{
\widetilde{\U}(h\mathbf e_a)
-
2\widetilde{\U}(\mathbf 0)
+
\widetilde{\U}(-h\mathbf e_a)
}{h^2},
\\
H_{ab}
&\simeq
\frac{
\widetilde{\U}[h(\mathbf e_a+\mathbf e_b)]
-
\widetilde{\U}[h(\mathbf e_a-\mathbf e_b)]
-
\widetilde{\U}[h(-\mathbf e_a+\mathbf e_b)]
+
\widetilde{\U}[-h(\mathbf e_a+\mathbf e_b)]
}{4h^2},
\qquad a\neq b.
\label{eq:fullspace-hessian}
\end{align}
A regular candidate is classified as a strict full-space local minimum when

\begin{equation}
    \frac{\|\mathbf g\|}{\pi \sum_i k_i}\leq\varepsilon_g
    \qquad\text{and}\qquad
    \lambda_{\min}(\mathbf H)>\varepsilon_{\lambda}.
\end{equation}

A stationary candidate with
\(\lambda_{\min}(\mathbf H)<-\varepsilon_{\lambda}\)
is classified as a saddle, while candidates within the eigenvalue tolerance are treated as marginal or numerically inconclusive. Configurations for which
\(\operatorname{rank}\mathbf C^{\star}<3\) are singular and are not classified by this local Hessian test.

For diagnostic purposes, the tangent space is further decomposed into symmetry-preserving and symmetry-breaking perturbations. Let
\(\mathbf A\delta\boldsymbol{\varphi}=\mathbf 0\)
denote the linearized fold-angle equalities associated with the imposed symmetry. If the columns of \(\mathbf Q_{\mathrm{sym}}\) span

$$
    \ker(\mathbf A\mathbf Z),
$$

and \(\mathbf Q_{\mathrm{br}}\) spans its orthogonal complement, then

\begin{equation}
    \mathbf H_{\mathrm{br}}
    =
    \mathbf Q_{\mathrm{br}}^{T}
    \mathbf H
    \mathbf Q_{\mathrm{br}}
\end{equation}

is the symmetry-breaking Hessian block. A negative eigenvalue of this block identifies a symmetry-breaking direction of negative curvature. For the regular degree-6 and degree-8 vertices considered here, the full tangent-space dimensions are three and five, respectively; the corresponding symmetry-breaking subspaces have dimensions one and three. Stability is nevertheless classified using the eigenvalues of the complete Hessian \(\mathbf H\), including any coupling between symmetry-preserving and symmetry-breaking coordinates.

The numerical procedure for each symmetry-reduced candidate is therefore:
\begin{enumerate}
\item reconstruct the complete \(\NCreases\)-component fold-angle vector and verify nonlinear loop closure;
\item construct the deformed crease matrix \(\mathbf C^{\star}\) and its tangent and normal bases \(\mathbf Z\) and \(\mathbf Y\);
\item retract every finite-difference perturbation onto exact loop closure using \eqref{eq:fullspace-retraction};
\item evaluate \(\mathbf g\), \(\mathbf H\), and the symmetry-breaking diagnostics; and
\item repeat the calculation at several finite-difference step sizes and retain a strict-minimum classification only when it is insensitive to this choice.
\end{enumerate}
With the stiffness normalization used in Section~3, the calculations use

\begin{equation}
    h\in
    \left\{
        3\times10^{-4},
        10^{-3},
        3\times10^{-3}
    \right\},
    \qquad
    \varepsilon_g
    =
    5\times10^{-3},
    \varepsilon_{\lambda}
    =
    10^{-6}.
\end{equation}

The nonlinear retraction is solved with a closure tolerance of \(10^{-9}\), and the resulting state is accepted only when the residual norm is no greater than \(10^{-9}\). A candidate is counted as a robust full-space minimum only when all three finite-difference step sizes give the strict-local-minimum classification.

Self-contact constraints are omitted from this auxiliary calculation. The test therefore examines stability in an enlarged compatible space in which facet penetration is permitted. Any candidate that remains a strict minimum in this enlarged space is necessarily stable when nonpenetration constraints are restored. Conversely, a descending direction found by this test may be blocked by self-contact and therefore does not, by itself, establish physical instability. Accordingly, for the symmetry-reduced candidates considered here,
\begin{equation}
N_{\mathrm{relaxed}}
\leq
N_{\mathrm{physical}}
\leq
N_{\mathrm{sym}},
\label{eq:fullspace-stability-bounds}
\end{equation}
where \(N_{\mathrm{relaxed}}\) is the number of candidates surviving the contact-relaxed full-space test and \(N_{\mathrm{sym}}\) is the number of minima identified in the contact-admissible symmetry-reduced landscape. Source code, candidate-state data, and the associated numerical diagnostics are provided in the github repository, \url{https://github.com/grasingerm/RSPA-2026-0429}.

}

\section{Symmetric cone inversion and area hiding} \label{app:cone}

We consider here a circular continuum membrane with radius $\Rad$ deformed into an axisymetric cone.
The membrane is free to stretch areally, but each fiber from the tip of the cone to the outside radius is constrained to constant length (\fref{fig:cone-analogy}.b).
In this case, the surface of the cone can be parameterized by
\begin{equation}
\xC = \left(\rad \cos \azi \sin \elevAngle, \rad \sin \azi \sin \elevAngle, \left(\Rad - \rad\right) \cos \elevAngle\right),
\end{equation}
where $\rad \in \left[0, \Rad\right], \azi \in [0, 2\pi)$.
Basis vectors in the tangent space are obtained by
\begin{align*}
\genBasis_{\rad} &= \frac{\partial \xC}{\partial \rad} = \left(\cos \azi \sin \elevAngle, \sin \azi \sin \elevAngle, -\cos \elevAngle\right), \\
\genBasis_{\azi} &= \frac{\partial \xC}{\partial \azi} = \left(-\rad \sin \azi \sin \elevAngle, \rad \cos \azi \sin \elevAngle, 0\right),
\end{align*}
and the metric tensor as
\begin{equation}
\metric_{\alpha \beta} = \genBasis_{\alpha} \cdot \genBasis_{\beta}.
\end{equation}
The area of a differential patch in the deformed configuration is
\begin{equation}
\df{a} = \sqrt{\det \metricTens} \: \df{\rad}\df{\azi} = \rad \left|\sin \elevAngle\right| \: \df{\rad}\df{\azi}
\end{equation}
so that the areal stretch is given by
\begin{equation} \label{eq:cone-areal-stretch}
\arealStretch \coloneqq \frac{\df{a}}{\df{A}} = \left|\sin \elevAngle\right| = \sin \left|\elevAngle\right|.
\end{equation}

The areal stretch of the continuum cone has an instructive relationship with the area hidden by folding of the origami cone.
Let $A_i$ be the area of facet $i$.
We define the ``visible area'', $A_i'$, of a facet as its area projected onto the surface of its analog continuum cone.
For the regular waterbomb folding of the origami cone,
\begin{equation}
	A_i' = A_i \left|\sin \frac{\foldAngle_i}{2}\right| = A_i \sin \left|\frac{\foldAngle_i}{2}\right|.
\end{equation}
As mentioned in \Fref{sec:origami-cone}, in the limit of an infinite number of creases
\begin{equation}
	\lim_{N \rightarrow \infty} \left|\frac{\foldAngle_i}{2}\right| = \left|\elevAngle\right|,
\end{equation}
and, consequently, the ratio of the visible area of a facet to its absolute area, $A_i' / A_i$ corresponds exactly with the areal stretch of the continuum cone given by \eqref{eq:cone-areal-stretch}.

\bibliographystyle{unsrtnat}
\bibliography{master}

\end{document}